\documentclass[transaction]{IEEEtran}
\usepackage{graphicx}
\usepackage{epstopdf}
\usepackage[latin1]{inputenc}
\usepackage{amsmath,amssymb,amscd,latexsym,dsfont}
\usepackage[english]{babel}
\usepackage{tabularx,cite}
\usepackage{graphicx}
\usepackage{algpseudocode}
\usepackage{algorithm}
\usepackage{algorithmicx}
\usepackage{float}
\usepackage{multicol}
\usepackage{psfrag}
\usepackage{comment,psfrag,subfigure,enumerate}
\usepackage{color}
\usepackage{amsthm}
\usepackage{graphicx}
\usepackage{graphicx}
\usepackage{epstopdf}
\usepackage[latin1]{inputenc}
\usepackage{amsmath,amssymb,amscd,latexsym,dsfont}
\usepackage[english]{babel}
\usepackage{tabularx,cite}
\usepackage{graphicx}
\usepackage{algpseudocode}
\usepackage{algorithm}
\usepackage{algorithmicx}
\usepackage{float}
\usepackage{multicol}
\usepackage{mathtools}
\usepackage{psfrag}
\usepackage{comment,psfrag,subfigure,enumerate}
\usepackage{color}
\usepackage{amsthm}
\ifCLASSINFOpdf
\else
\fi
\begin{document}
\IEEEoverridecommandlockouts
\IEEEpubid{\makebox[\columnwidth]{978-1-4799-5863-4/14/\$31.00 \copyright 2014 IEEE \hfill} \hspace{\columnsep}\makebox[\columnwidth]{ }}
\title{On the Performance of RF-FSO Links with and without Hybrid ARQ}

\author{\IEEEauthorblockN{Behrooz Makki, Tommy Svensson,  \emph{Senior Member, IEEE}, Thomas Eriksson and Mohamed-Slim Alouini, \emph{Fellow, IEEE}}\\
\thanks{Behrooz Makki, Tommy Svensson and Thomas Eriksson are with Chalmers University of Technology, Email: \{behrooz.makki, tommy.svensson, thomase\}@chalmers.se. Mohamed-Slim Alouini is with the King Abdullah University of Science and Technology (KAUST), Email: slim.alouini@kaust.edu.sa}
\thanks{Part of this work has been accepted for presentation at the IEEE GLOBECOM 2015.}
}

%


\maketitle
\vspace{-15mm}
\begin{abstract}
This paper studies the performance of hybrid radio-frequency (RF) and free-space optical (FSO) links assuming perfect channel state information (CSI) at the receiver. Considering the cases with and without hybrid automatic repeat request (HARQ), we derive closed-form expressions for the message decoding probabilities as well as the throughput and the outage probability of the RF-FSO setups. We also evaluate the effect of adaptive power allocation and different channel conditions on the throughput and the outage probability. The results show the efficiency of the RF-FSO links in different conditions.
\end{abstract}



%
\IEEEpeerreviewmaketitle
\vspace{-0mm}
\section{Introduction}
The next generation of communication networks must provide high-rate reliable data streams. To address these demands, a combination of different techniques are considered among which free-space optical (FSO) communication is very promising  \cite{6331134,6932439,6887284}. FSO systems provide fiber-like data rates through the atmosphere using lasers or light emitting diodes (LEDs). Thus, the FSO can be used for a wide range of applications such as last-mile access, fiber back-up, back-haul for wireless cellular networks, and disaster recovery. However, such links are highly susceptible to atmospheric effects and, consequently, are unreliable. An efficient method to improve the reliability in FSO systems is to rely on an additional radio-frequency (RF) link to create a hybrid RF-FSO communication system.

Typically, to achieve data rates comparable to those in the FSO link, a millimeter wavelength carrier is selected for the RF link. As a result, the RF link is also subject to atmospheric effects such as rain and scintillation. However, the good point is that these links are complementary because the RF (resp. the FSO) signal is severely attenuated by rain (resp. fog/clouds) while the FSO (resp. the RF) signal is not. Therefore, the link reliability and the service availability are considerably improved via joint RF-FSO based data transmission.

The performance of RF-FSO systems is studied in different papers, e.g., \cite{6887284,4610745,1399401,4168193,4393998,Hamzeh,6364576,6400459}, where the RF and the FSO links are considered as separate links and the RF link acts as a backup when the FSO link is down. In the meantime, there are works such as \cite{6503564,5342330,4411336,4348339,5351671,5427418} in which the RF and the FSO links are combined to improve the system performance. Also, \cite{6831655,6866170} study RF-FSO based relaying schemes with an RF source-relay link and an FSO \cite{6831655,6866170} or RF-FSO \cite{6866170} relay-destination link. Moreover, the implementation of hybrid automatic repeat request (HARQ) in RF-based (resp. FSO-based) systems is investigated in, e.g., \cite{outageHARQ,Tcomkhodemun,throughputdef,a01661837,MIMOARQkhodemun,tuninetti2011,6006606,letterzorzikhodemun} (resp. \cite{6168189,6139812,6857378,5380093,5557647,5680280}), while the HARQ-based RF-FSO systems have been rarely studied, e.g., \cite{5427418,6692504}.

In this paper, we study the data transmission efficiency of RF-FSO systems from an information theoretic point of view. We derive closed-form expressions for the message decoding probabilities as well as the system throughput and outage probability (Lemmas 1-5, Eq. (\ref{eq:equpperboundG}), (\ref{eq:eqlowermolla}), (\ref{eq:eqxxxx})). The results are obtained in the cases with and without HARQ. Also, we analyze the effect of adaptive power allocation between the RF and the FSO links on the throughput/outage probability (Lemmas 6-7 and Fig. 11). Finally, we investigate the effect of different channel conditions on the performance of RF-FSO setups and compare the results with the cases utilizing either the RF or the FSO link separately. Note that, while the results are presented for the RF-FSO setups, with proper refinements of the channel model, the same analysis as in the paper is useful for other coordinated data transmission schemes as well (see Section III.C for more discussions).

As opposed to \cite{6887284,4610745,1399401,4168193,4393998,Hamzeh,6364576,6400459}, we consider joint data transmission/reception in the RF and FSO links. Also, the paper is different from \cite{6503564,5342330,4411336,6831655,6866170,4348339,5351671,5427418,Tcomkhodemun,throughputdef,outageHARQ,a01661837,MIMOARQkhodemun,tuninetti2011,6006606,letterzorzikhodemun,5557647,6168189,6139812,6857378,5380093,5680280,6692504} because we study the performance of HARQ in joint RF-FSO links and derive new analytical/numerical results on the message decoding probabilities, optimal power allocation, and outage probability/throughput which have not been presented before. The derived results provide a framework for the analysis of RF-FSO links from different perspectives.

The numerical and the analytical results show that 1) depending on the relative coherence times of the RF and FSO links there are different methods for the analytical performance evaluation of the RF-FSO systems (Lemmas 1-5, Eq. (\ref{eq:equpperboundG}), (\ref{eq:eqlowermolla}), (\ref{eq:eqxxxx})). 2) While adaptive power allocation improves the system performance, at high signal-to-noise ratios (SNRs), the optimal, in terms of throughput/outage probability, power allocation between the RF and the FSO links converges to uniform power allocation, independently of the links channel conditions (Lemma 6). Finally, 3) the joint implementation of the RF and FSO links leads to substantial performance improvement, compared to the cases with only the RF or the FSO link. For instance, consider the exponential distribution and the common relative coherence times of the RF and FSO links. Then, with the initial code rate $R=5$ nats-per-channel-use (npcu), a maximum of $M=2$ retransmission rounds of the HARQ and the outage probability $10^{-2}$, the joint RF-FSO based data transmission reduces the required power by $16$ and $4$ dB, compared to the cases with only the RF or the FSO link, respectively (see Fig. 10 for more details).
\vspace{-0mm}
\section{System Model and Problem Formulation}
\begin{figure}
\centering
  \includegraphics[width=0.98\columnwidth]{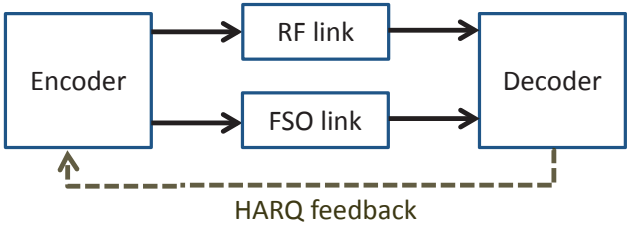}\\\vspace{-2mm}
\caption{Channel model. The data is jointly transmitted by the RF and the FSO links and, in each round of HARQ, the receiver decodes the data based on all received signals.}\vspace{-2mm}\label{figure111}
\end{figure}
\vspace{-0mm}
\subsection{System Model}
Consider a joint RF-FSO system, as demonstrated in Fig. 1. The data sequence is encoded
into parallel FSO and RF bit streams. The FSO link employs intensity modulation and direct detection while the RF link modulates the encoded bits and up-converts the baseband signal to a millimeter wavelength, in the range of $30-300$ GHz, RF carrier frequency\footnote{A millimeter wavelength carrier is often selected for the RF link to achieve data rates comparable to those in the FSO link. However, this is not a necessary assumption, and the theoretical results hold for different ranges of carrier frequencies.}. Then, the FSO and the RF signals are simultaneously sent to the receiver. At the receiver, the received RF (FSO) signal is
down-converted to baseband (resp. collected by an aperture and converted to an electrical signal via photo-detection) and the signals are sent to the decoder which decodes the received signals jointly (also, see \cite{6503564,5342330,4411336,4348339,5351671,5427418} for further discussions on the coding process in RF-FSO setups). In this way, as seen in the following, the diversity increases by the joint data transmission of the RF and FSO links, and one link can compensate the poor performance of the other link experiencing severe atmospheric effects.

We denote the instantaneous realizations of the fading coefficient of the RF link and the turbulence coefficient of the FSO link in time slot $i$ by $H_{\text{RF},i}$ and $H_{\text{FSO},i}$, respectively, and for simplicity we refer to both of them as channel coefficients. These channel coefficients are assumed to be known at the receiver which is an acceptable assumption in block-fading conditions  \cite{Tcomkhodemun,throughputdef,outageHARQ,a01661837,MIMOARQkhodemun,tuninetti2011,6006606,letterzorzikhodemun,6168189,6139812,6857378}. Also, we define $G_{\text{RF},i}=|H_{\text{RF},i}|^2,$ $G_{\text{FSO},i}=|H_{\text{FSO},i}|^2$ which are referred to as channel gain realizations in the following. We then assume no channel state information (CSI) feedback to the transmitter, except for the HARQ feedback bits. In each round of data transmission, if the data is correctly decoded by the receiver, an acknowledgement (ACK) is fed back to the transmitter, and the retransmissions stop. Otherwise, the receiver transmits a negative-acknowledgment (NACK). The feedback channel can be an RF, an FSO or an RF-FSO link, and is supposed to be error- and delay-free. However, we can follow the same procedure as in \cite{letterzorzikhodemun} and \cite{noisyARQkhodemun} to extend the results to the cases with delayed and erroneous feedback, respectively. The results are presented for asymptotically long codes where the block error rate converges to the outage probability. Finally, we assume perfect synchronization between the links which is an acceptable assumption in our setup with the RF and the FSO signals generated at the same transmit terminal, e.g., \cite{6503564,5342330,4411336,4348339,5351671,5427418,6692504} (to study practical schemes for synchronization between the RF and FSO links, the reader is referred to \cite{6866170}).

Let us define a packet as the transmission of a codeword along with all its possible retransmissions. As the most promising HARQ approach leading to highest throughput/lowest outage probability \cite{Tcomkhodemun,throughputdef,MIMOARQkhodemun,a01661837,tuninetti2011,6006606,letterzorzikhodemun}, we consider the incremental redundancy (INR) HARQ with a maximum of $M$ retransmissions, i.e., the message is retransmitted a maximum of $M$ times. Using INR HARQ, $K$ information nats are encoded into a \emph{parent} codeword of length $ML$ channel uses. The parent codeword is then divided into $M$ sub-codewords of length $L$ channel uses which are sent in the successive transmission rounds. Thus, the equivalent data rate, i.e., the code rate, at the end of round $m$ is $\frac{K}{mL}=\frac{R}{m}$ where $R=\frac{K}{L}$ denotes the initial code rate. In each round, the receiver combines all received sub-codewords to decode the message. The retransmission continues until the message is correctly decoded or the maximum permitted transmission round is reached. Note that setting $M=1$ represents the cases without HARQ, i.e., open-loop communication.

Finally, the results are valid for different ranges of RF and FSO links operation frequencies. Thus, for generality and in harmony with, e.g., \cite{Tcomkhodemun,throughputdef,a01661837,MIMOARQkhodemun,tuninetti2011,6006606,letterzorzikhodemun}, we do not specify the frequency ranges of the RF and FSO links and present the code rates in the nats-per-channel-use (npcu) units. The results of the simulation figures can be easily mapped to the bit-per-channel-use scale if the code rates are scaled by $\log_2e$.  Moreover, the results are represented in nats-per-second/Hertz (npsH), if each channel use is associated with a time-frequency unit. Also, for given bandwidths of the RF and FSO links, we can follow the same approach as in \cite[Chapter 9.3]{4444444444} to present the results in bits- or nats-per-second/Hertz.
\vspace{-4mm}
\subsection{Problem Formulation}
It has been previously showed that for different channel models, the throughput and the outage probability of different HARQ protocols can be written as \cite{throughputdef,a01661837,MIMOARQkhodemun,tuninetti2011}
\begin{align}\label{eq:eqeta1}
\eta=R\frac{1-\Pr(W_M\le \frac{R}{M})}{1+\sum_{m=1}^{M-1}{\Pr(W_m\le \frac{R}{m})}}
\end{align}
and
\vspace{-3mm}
\begin{align}\label{eq:eqoutprob1}
\Pr(\text{Outage})=\Pr\left(W_M\le \frac{R}{M}\right),
\end{align}
%
respectively, where $W_m$ is the accumulated mutual information (AMI) at the end of round $m$. Also, $\Pr(W_m\le \frac{R}{m})$ denotes the probability that the data is not correctly decoded up to the end of the $m$-th round.  In this way, the throughput and the outage probability of HARQ protocols are monotonic functions of the probabilities $\Pr(W_m\le \frac{R}{m}),\forall m$. This is because the system performance depends on the retransmission round in which the codewords are correctly decoded. Moreover, the probability $\Pr(W_m\le \frac{R}{m})$ is directly linked to the AMI $W_{m}$ which is a random variable and function of the channel realizations experienced in rounds $n=1,\ldots,m.$ As such, to analyze the throughput and the outage probability, the key point is to determine the AMIs as functions of channel realization(s) and find their corresponding cumulative distribution functions (CDFs)\footnote{The CDF and the probability distribution function (PDF) of a random variable $X$ are denoted by $F_X(.)$ and $f_X(.),$ respectively.} $F_{W_{m}},m=1,\ldots,M$. Then, having the CDFs, the probabilities $\Pr(W_m\le \frac{R}{m})$ and, consequently, the considered performance metrics are obtained. Therefore, instead of concentrating on (\ref{eq:eqeta1})-(\ref{eq:eqoutprob1}), we first find the CDFs $F_{W_{m}},m=1,\ldots,M,$ for the INR-based RF-FSO system.

To find the CDFs, we utilize the properties of the RF and the FSO links to derive the AMIs as in (\ref{eq:eqWm1}). Since there is no closed-form expression for the CDFs of the AMIs, we need to use different approximation techniques. In the first method, we use the central limit Theorem (CLT) to approximate the contribution of the FSO link on the AMI by an equivalent Gaussian random variable. Using the CLT, we find the mean and the variance of the equivalent random variable for the exponential, log-normal and Gamma-Gamma distributions of the FSO link as given in (\ref{eq:mueq})-(\ref{eq:sigmaeq}), (\ref{eq:eqmulognormal})-(\ref{eq:eqsigmalognormal}) and (\ref{eq:eqmugamma1})-(\ref{eq:eqsigmagamma1}), respectively. With the derived means and variances of the Gaussian variable, we find the CDF of the AMIs in Lemmas 1-5 (see Section III.A for details). An alternative approximation approach, presented in Section III.B,  is to approximate the PDF of the AMI in the FSO link and then find the PDF of the joint RF-FSO setup, as given in (\ref{eq:equpperboundG}), (\ref{eq:eqlowermolla}), (\ref{eq:eqxxxx}). Finally, we use the properties of the AMIs to analyze the optimal power allocation between the RF and the FSO  links as given in Lemmas 6-7. In Section IV, we validate the accuracy of the approximations and evaluate the throughput/outage probability of the RF-FSO system for different channel conditions.
\vspace{-0mm}
\section{Analytical Results}
\begin{figure}
\centering
  \includegraphics[width=0.98\columnwidth]{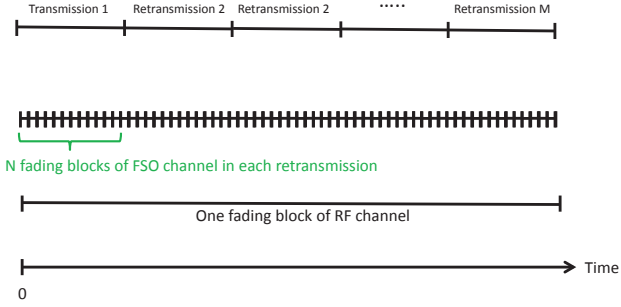}\\\vspace{-2mm}
\caption{Time scales. The RF link is supposed to remain constant in the retransmissions (quasi-static channel \cite{Tcomkhodemun,a01661837,MIMOARQkhodemun,6006606,letterzorzikhodemun}) while in each retransmission round of HARQ $N$ different channel realizations are experienced in the FSO link. In Section III.A, we study the cases with large values of $N$. Section III.B analyzes the system performance for the small values of $N$, i.e., when the coherence times of the RF and FSO links are of the same order.}\vspace{-2mm}\label{figure111}
\end{figure}

In RF-FSO systems, it was demonstrated by, e.g., \cite{5342330,1142964,598416,598420}, that the RF link experiences very slow variations and the coherence time of the RF link is in the order of $10^{2}-10^{3}$ times larger than the coherence time of the FSO link. Here, we start the discussions by considering the setup as illustrated in Fig. 2 where the RF link remains constant in the retransmissions (quasi-static channel \cite{Tcomkhodemun,a01661837,MIMOARQkhodemun,6006606,letterzorzikhodemun}) while in each retransmission round of HARQ $N$ different channel realizations are experienced in the FSO link. However, note that this is not a necessary condition because 1) the same analysis holds for the cases with shorter coherence time of the RF link, compared to the coherence time of the FSO link and 2) as seen in Section III.B, we can derive the results in the cases with few, possibly 1, channel realizations of the FSO link during the packet transmission.

Considering Fig. 2, we can use the results of \cite[Chapter 7]{ELGAMAL} and \cite[Chapter 15]{4444444444} to find the AMI of the  joint RF-FSO link at the end of the $m$-th round as
\vspace{-1mm}
\begin{align}\label{eq:eqWm1}
&W_m=\log(1+P_\text{RF}G_\text{RF})\nonumber\\&\,\,\,\,\,\,\,\,\,\,\,+\frac{\psi}{m}\sum_{j=1}^{m}{\left(\frac{1}{N}\sum_{i=1}^N{\log(1+cP_\text{FSO}G_{\text{FSO},1+(j-1)i})}\right)}
\nonumber\\&\,\,\,\,\,\,\,\,\,\,\,=\log(1+P_\text{RF}G_\text{RF})+\mathcal{Y}_{(m,N)},
\nonumber\\&\mathcal{Y}_{(m,N)}\doteq\frac{\psi}{mN}\sum_{j=1}^{m}{\sum_{i=1}^N{\log(1+cP_\text{FSO}G_{\text{FSO}, 1+(j-1)i})}}.
\end{align}
Here, $P_\text{RF}$ and $P_\text{FSO}$ are, respectively, the transmission powers in the RF and FSO links. Also, $G_\text{RF}$ and $G_{\text{FSO},j}$'s denote the channel gain realizations of the RF and the FSO links in different retransmission rounds, respectively. Then, $\psi$ represents the relative symbol rates of the RF and FSO links which is a design parameter selected by the network designer. With no loss of generality, we set $\psi=1$ in the following, while the results can be easily extended to the cases with different values of $\psi$. Also, (\ref{eq:eqWm1}) is based on the fact that the achievable rate of an FSO link is given by $\log(1+cx)$ with $x$ being the instantaneous received SNR and $c$ denoting a constant term such that $c = 1$ for heterodyne detection and $c =\frac{e}{2\pi}$ for intensity modulation/direct detection (IM/DD) \cite{7192727}, \cite[Eq. 26]{5238736}, \cite[Eq. 7.43]{fsobookalouini}, \cite{alouinicapacityakhar}\footnote{In \cite{5238736}, $\log(1+c\text{SNR})$ is proved as a tight lower bound on the capacity in the cases with an average power constraint. Then, \cite{7192727,alouinicapacityakhar} show that the formula of the kind $\log(1+c\text{SNR})$ is an asymptotically tight lower bound on the achievable rates for the cases with an average power constraint, a peak power constraint, as well as combined peak and average power constraints.}. In the following, we set $c=1$ which corresponds to the cases with heterodyne detection. In the meantime, setting $c =\frac{e}{2\pi}$, it is straightforward to extend the results to the cases with IM/DD. Finally, as the noise variances at the receiver are set to 1, we define $10\log_{10}P, P=P_\text{RF}+P_\text{FSO},$ as the SNR.
\vspace{-0mm}
\subsection{Performance Analysis in the Cases with Considerably Different Coherence Times for the RF and FSO Links}
Here, we consider the cases where the coherence times of the RF and FSO links are considerably different. Motivated by, e.g., \cite{5342330,1142964,598416,598420}, we concentrate on the cases with shorter coherence time of the FSO link, compared to the RF link. Meanwhile, the same analysis is valid, if the RF link experiences shorter coherence time than the FSO link.

Considering the conventional channel conditions of the RF and FSO links and different values of $N$, there is no closed-form expression for the CDF of $W_m,\forall m.$ Therefore, we use the CLT to approximate $\mathcal{Y}_{(m,N)}$ by the Gaussian random variable $\mathcal{Z}\sim\mathcal{N}(\mu,\frac{1}{mN}\sigma^2)$ where $\mu$ and $\sigma^2$ are the mean and variance determined based on the FSO link channel condition.

Reviewing the literature and depending on the channel condition, the FSO link is commonly considered to follow exponential, log-normal or Gamma-Gamma distributions, e.g., \cite{5342330,6168189,FSObook}. For the exponential distribution of the FSO link, i.e., $f_{G_\text{FSO}}(x)=\lambda e^{-\lambda x}$ with $\lambda$ being the long-term channel coefficient, we have
\vspace{-0mm}
\begin{align}\label{eq:mueq}
\mu&=E\{\log(1+P_\text{FSO}G_\text{FSO})\}=\int_0^\infty{f_{{G_{\text{FSO}}}}(x)\log(1+P_\text{FSO}x)\text{d}x}\nonumber\\&\mathop  = \limits^{(a)} P_\text{FSO}\int_0^\infty{\frac{1-F_{G_{\text{FSO}}}(x)}{1+P_\text{FSO}x}\text{d}x}=-e^{\frac{\lambda }{P_\text{FSO}}}\text{Ei}\left(-{\frac{\lambda }{P_\text{FSO}}}\right)
\end{align}
and
\vspace{-0mm}
\begin{align}\label{eq:sigmaeq}
&\sigma^2=\rho^2-\mu^2,\nonumber\\&
\rho^2=E\{\log(1+P_\text{FSO}G_\text{FSO})^2\}\nonumber\\&=\int_0^\infty{f_{{G_{\text{FSO}}}}(x)\log^2(1+P_\text{FSO}x)\text{d}x} \nonumber\\& \mathop  = \limits^{(b)}{2P_\text{FSO}}\int_0^\infty{\frac{e^{- \lambda x}}{1+P_\text{FSO}x}\log(1+P_\text{FSO}x)\text{d}x}\mathop  \simeq \limits^{(c)} \mathcal{K}\left(\infty\right)-\mathcal{K}\left(1\right),\nonumber\\&
\mathcal{K}(x)=\frac{2}{\lambda}e^{\frac{\lambda}{P_\text{FSO}}}\bigg(\frac{\lambda}{P_\text{FSO}}x\prescript{}{3}F_{3}\left(1,1,1;2,2,2;-\frac{\lambda x}{P_\text{FSO}}\right)\nonumber\\&+\frac{1}{2}\log(x)\bigg(-2\left(\log\left(\frac{\lambda}{P_\text{FSO}}x\right)+\mathcal{E}\right)\nonumber\\&-2\Gamma\left(0,\frac{\lambda}{P_\text{FSO}}x\right)+\log(x)\bigg)\bigg).
\end{align}
Here, $E\{.\}$ denotes the expectation operator. Also, $(a)$ and $(b)$ are obtained by partial integration. Then, denoting the Euler constant by $\mathcal{E}$, $(c)$ is given by variable transform $1+P_\text{FSO}x=t,$ some manipulations and the definition of Gamma incomplete function $\Gamma(s,x)=\int_x^\infty{t^{s-1}e^{-t}\text{d}t}$ and the generalized hypergeometric function $\prescript{}{a_1}F_{a_2}(.).$

For the log-normal distribution of the FSO link, i.e., $f_{G_\text{FSO}}(x)=\frac{1}{\sqrt{2\pi}{\delta x}}e^{-\frac{(\log(x)-\varpi)^2}{2\delta^2}}$ where $\delta$ and $\varpi$ represent the long-term channel parameters, the mean $\mu$ is rephrased as

\begin{align}
\mu&= P_\text{FSO}\int_0^\infty{\frac{1-F_{G_{\text{FSO}}}(x)}{1+P_\text{FSO}x}\text{d}x} \nonumber\\& \mathop  = \limits^{(d)}\frac{P_\text{FSO}}{2}\int_0^\infty{\frac{1-\text{erf}\left(\frac{\log(x)-\varpi}{\sqrt{2}\delta}\right)}{1+P_\text{FSO}x}\text{d}x}\nonumber\\&=P_\text{FSO}\int_0^\infty{\frac{Q\left(\frac{\log(x)-\varpi}{\delta}\right)}{1+P_\text{FSO}x}\text{d}x} \nonumber\\& \mathop  \simeq \limits^{(e)}
P_\text{FSO}\int_0^\infty{\frac{U(x)}{1+P_\text{FSO}x}\text{d}x}
\nonumber\\&=P_\text{FSO}\bigg(\int_0^{\max\left(0,\frac{1}{2c}+e^{\varpi}\right)}{\frac{1}{1+P_\text{FSO}x}\text{d}x}\nonumber\\&\,\,\,\,\,\,\,\,\,\,\,\,\,\,\,\,\,\,\,\,\,\,\,\,+\int_{\max\left(0,\frac{1}{2c}+e^{\varpi}\right)}^{\frac{-1}{2c}+e^{\varpi}}{\frac{\frac{1}{2}+c(x-e^\varpi)}{1+P_\text{FSO}x}\text{d}x}\bigg)\nonumber
\end{align}
\begin{align}\label{eq:eqmulognormal}
&=\log\left(1+P_\text{FSO}\max\left(0,\frac{1}{2c}+e^{\varpi}\right)\right)\nonumber\\&+\left(\frac{1}{2}-ce^\varpi-\frac{c}{P_\text{FSO}}\right)\log\left(\frac{1+P_\text{FSO}\left(\frac{-1}{2c}+e^{\varpi}\right)}{1+P_\text{FSO}\max\left(0,\frac{1}{2c}+e^{\varpi}\right)}\right)\nonumber\\&+{c}\left(\frac{-1}{2c}+e^{\varpi}-\max\left(0,\frac{1}{2c}+e^{\varpi}\right)\right), c\doteq \frac{-e^{-\varpi}}{\delta\sqrt{2\pi}}.
\end{align}
Here, $\text{erf}(x)=\frac{2}{\sqrt{\pi}}\int_0^x{e^{-t^2}\text{d}t}$ and $Q(x)=\frac{1}{\sqrt{2\pi}}\int_x^\infty{e^{-\frac{t^2}{2}}\text{d}t}$ represent the error and the Gaussian $Q$ functions, respectively. Moreover, $(d)$ holds for the log-normal distribution and $(e)$ comes from the linearization technique
\vspace{-0mm}
\begin{align}\label{eq:linearizationnik}
&Q\left(\frac{\log(x)-\varpi}{\delta}\right)\simeq U(x), \nonumber\\&
U(x)= \left\{\begin{matrix}
1 & x< \frac{1}{2c}+e^{\varpi}, \\
\frac{1}{2}+c(x-e^\varpi) & x\in[\frac{1}{2c}+e^{\varpi},\frac{-1}{2c}+e^{\varpi}],\\
0 & x> \frac{-1}{2c}+e^{\varpi},
\end{matrix}\right.
\end{align}
with
\vspace{-2mm}
\begin{align}\label{eq:eqformulC}
c=\frac{\partial \left(Q\left(\frac{\log(x)-\varpi)}{\sqrt{2}\delta}\right)\right)}{\partial x}\bigg|_{x=e^\varpi}=\frac{-e^{-\varpi}}{\delta\sqrt{2\pi}},
\end{align}
which is found by the derivative of $Q\left(\frac{\log(x)-\varpi}{\sqrt{2}\delta}\right)$ at point $x=e^\varpi.$
Also, following the same procedure as in (\ref{eq:eqmulognormal}), the variance $\sigma^2$ is determined as
\begin{align}\label{eq:eqsigmalognormal}
&\sigma^2=\rho^2-\mu^2,\nonumber\\&
\rho^2=2P_\text{FSO}\int_0^\infty{\frac{\log(1+P_\text{FSO}x)}{1+P_\text{FSO}x}Q\left(\frac{\log(x)-\varpi}{\delta}\right)\text{d}x}\nonumber\\&\mathop  \simeq \limits^{(f)} 2P_\text{FSO}\int_0^\infty{\frac{\log(1+P_\text{FSO}x)}{1+P_\text{FSO}x}U(x)\text{d}x}\nonumber\\&
=\left(\log\left(1+P_\text{FSO}\max\left(0,\frac{1}{2c}+e^{\varpi}\right)\right)\right)^2+\nonumber\\&
\frac{1}{P_\text{FSO}}\Bigg(     \Big(({\frac{1}{2}-ce^\varpi})P_\text{FSO}-c\Big)\times\nonumber\\&\bigg(\Big(\log\Big(1+P_\text{FSO}\big(\frac{-1}{2c}+e^{\varpi}\big)\Big)\Big)^2\nonumber\\&-\Big(\log\Big(1+P_\text{FSO}\max\big(0,\frac{1}{2c}+e^{\varpi}\big)\Big)\Big)^2\bigg) \nonumber\\& -2cP_\text{FSO}\left(\frac{-1}{2c}+e^{\varpi}-\max\left(0,\frac{1}{2c}+e^{\varpi}\right)\right)
\nonumber\\&+2c\Bigg(\Big(1+P_\text{FSO}(\frac{-1}{2c}+e^{\varpi})\Big)\log\Big(1+P_\text{FSO}\big(\frac{-1}{2c}+e^{\varpi}\big)\Big)\nonumber\\&-\left(1+P_\text{FSO}\max\Big(0,\frac{1}{2c}+e^{\varpi}\Big)\right)\times\nonumber\\&\log\left(1+P_\text{FSO}\max\Big(0,\frac{1}{2c}+e^{\varpi}\Big)\right)\Bigg)                 \Bigg),
\end{align}
for the log-normal distribution of the FSO link, where $(f)$ comes from (\ref{eq:linearizationnik})-(\ref{eq:eqformulC}).

Finally, we consider the Gamma-Gamma distribution for the FSO link in which the channel gain follows
\begin{align}\label{eq:eqpdfgammagamma}
f_{G_\text{FSO}}(x)=\frac{2(ab)^{\frac{a+b}{2}}}{\Gamma(a)\Gamma(b)}x^{\frac{a+b}{2}-1}K_{a-b}\left(2\sqrt{abx}\right),
\end{align}
with $K_n$ denoting the modified Bessel function of the second kind of order $n$ and $\Gamma(x)=\Gamma(x,0)$ being the Gamma function. Moreover, $a$ and $b$ are the distribution shaping parameters which can be expressed as functions of Rytov variance \cite{6168189,FSObook}.

For the Gamma-Gamma distribution, we can use (\ref{eq:eqpdfgammagamma}) and, e.g., the approximation schemes of (\ref{eq:gammagammaapx1})-(\ref{eq:gammagammaapx12}) to derive the mean and variance as expressed in equations (\ref{eq:eqmugamma1})-(\ref{eq:eqsigmagamma1}) of the Appendix.
Intuitively, (\ref{eq:mueq})-(\ref{eq:eqsigmalognormal}), (\ref{eq:eqmugamma1})-(\ref{eq:eqsigmagamma1}) indicate that, with considerably different coherence times of the RF and FSO links, there are mappings between the performance of RF-FSO systems with exponential, log-normal and Gamma-Gamma distributions of the FSO link, in the sense that with proper scaling of the channel parameters the same outage probability/throughput is achieved in these conditions.
Finally, note that $\lim_{N\to\infty}\frac{1}{mN}\sum_{j=1}^{m}{\sum_{i=1}^N{\log(1+P_\text{FSO}G_{\text{FSO}, 1+(j-1)i})}}= E\{\log(1+P_\text{FSO}G_{\text{FSO}})\}, \forall m.$ Intuitively, this means that for asymptotically large values of $N$, i.e., significantly shorter coherence time of the FSO link compared to the one in the RF link, the AMI of the FSO link converges to its ergodic capacity $E\{\log(1+P_\text{FSO}G_{\text{FSO}})\}$. Thus, in this case the RF-FSO link is mapped to an equivalent RF link in which successful decoding of the rate equal to the ergodic capacity of the FSO link is always guaranteed.

Having $\mu$ and $\sigma^2$, we find the CDFs $F_{W_{m}},\forall m,$ as follows. Consider Rayleigh-fading conditions for the RF link where the fading coefficients follow $H_\text{RF}\sim\mathcal{CN}(0,1)$ and, consequently, $f_{G_\text{RF}}(x)=e^{-x}, G_\text{RF}=|H_\text{RF}|^2.$ Using (\ref{eq:eqWm1}) and the mean and variance of $\mathcal{Z}$, the CDFs of the AMIs are given by
\begin{align}\label{eq:cdfWmapprox}
&F_{W_m}(u)=\Pr(\log(1+P_\text{RF}G_\text{RF})+\mathcal{Y}_{(m,N)}\le u)\nonumber\\&=\int_0^{\frac{e^u-1}{P_\text{RF}}}{f_{G_\text{RF}}(x)\Pr(\mathcal{Y}_{(m,N)}\le u-\log(1+P_\text{RF}G_\text{RF}))\text{d}x}\mathop  = \limits^{(g)}\nonumber\\&\int_0^{\frac{e^u-1}{P_\text{RF}}}{e^{-x}Q\left(\frac{\sqrt{mN}(\log(1+P_\text{RF}x)+\mu-u)}{\sigma}\right)\text{d}x}, m, N\ge1 ,
\end{align}
where $(g)$ comes from the CDF of Gaussian distributions and CLT. Also, for the exponential, log-normal and the Gamma-Gamma distribution of the FSO link the mean and variance $(\mu,\sigma^2)$ are given by (\ref{eq:mueq})-(\ref{eq:sigmaeq}), (\ref{eq:eqmulognormal})-(\ref{eq:eqsigmalognormal}) and (\ref{eq:eqmugamma1})-(\ref{eq:eqsigmagamma1}), respectively. Therefore, the final step to derive the throughput and the outage probability is to find (\ref{eq:cdfWmapprox}) while it does not have closed-form expression. The following lemmas propose several approximation/bounding approaches for the CDF of the AMIs and, consequently, the throughput/outage probability.

\emph{\textbf{Lemma 1}}: The throughput and the outage probability of the HARQ-based RF-FSO setup are approximately given by
\begin{align}\label{eq:eqapproxeta1}
\eta=R\frac{1-\mathcal{F}(\frac{R}{M})}{1+\sum_{m=1}^{M-1}{\mathcal{F}(\frac{R}{m})}}
\end{align}
and
\begin{align}\label{eq:eqapproxoutprob1}
\Pr(\text{Outage})=\mathcal{F}\left(\frac{R}{M}\right),
\end{align}
respectively, with $\mathcal{F}(x)$ defined in (\ref{eq:cdfWmapproxlemma}).
\begin{proof}
To find the approximations, we implement $Q\left(\frac{\sqrt{mN}(\log(1+P_\text{RF}x)+\mu-u)}{\sigma}\right)\simeq V(x)$ with
\begin{align}\label{eq:linearizationnik2}
&V(x)= \left\{\begin{matrix}
1 \,\,\,\,\,\,\,\,\,\,\,\,\,\,\,\,\,\,\, \text{if  } x< \frac{-\sigma\sqrt{\pi}e^{u-\mu}}{P_\text{RF}\sqrt{2mN}}+\frac{e^{u-\mu}-1}{P_\text{RF}},\,\,\,\,\,\,\,\,\,\,\,\,\,\,\,\,\,\,\,\,\,\,\,\,\,\,\,\,\,\,\,\,\, \\
\frac{1}{2}-\frac{\sqrt{mN}P_\text{RF}e^{\mu-u}\left(x-\frac{e^{u-\mu}-1}{P_\text{RF}}\right)}{\sigma\sqrt{2\pi}}\,\,\,\,\,\,\,\,\,\,\,\,\,\,\,\,\,\,\,\,\,\,\,\,\,\,\,\,\,\,\,\,\,\,\,\,\,\,\,\,\,\,\,\,\,\,\,\,\,\,\,\,\,\,\,\,\,\,\,\,\,\,\,\,\,\, \\
\,\,\,\,\,\,\,\,\text{if } x\in\Big[\frac{-\sigma\sqrt{\pi}e^{u-\mu}}{P_\text{RF}\sqrt{2mN}}+\frac{e^{u-\mu}-1}{P_\text{RF}},\frac{\sigma\sqrt{\pi}e^{u-\mu}}{P_\text{RF}\sqrt{2mN}}+\frac{e^{u-\mu}-1}{P_\text{RF}}\Big],&\\
0 \,\,\,\,\,\,\,\,\,\,\,\,\,\,\,\,\,\,\, \text{if  }  x> \frac{\sigma\sqrt{\pi}e^{u-\mu}}{P_\text{RF}\sqrt{2mN}}+\frac{e^{u-\mu}-1}{P_\text{RF}},\,\,\,\,\,\,\,\,\,\,\,\,\,\,\,\,\,\,\,\,\,\,\,\,\,\,\,\,\,\,\,\,\,
\end{matrix}\right.
\end{align}
which leads to
\begin{align}\label{eq:cdfWmapproxlemma}
&F_{W_m}(u)\simeq \int_0^{\frac{e^u-1}{P_\text{RF}}}{e^{-x}V(x)\text{d}x}\nonumber\\&= \int_0^{\max\left(0,\frac{-\sigma\sqrt{\pi}e^{u-\mu}}{P_\text{RF}\sqrt{2mN}}+\frac{e^{u-\mu}-1}{P_\text{RF}}\right)}{e^{-x}\text{d}x}\nonumber\\&+\int_{{\max\left(0,\frac{-\sigma\sqrt{\pi}e^{u-\mu}}{P_\text{RF}\sqrt{2mN}}+\frac{e^{u-\mu}-1}{P_\text{RF}}\right)}}^{\min\left(\frac{\sigma\sqrt{\pi}e^{u-\mu}}{P_\text{RF}\sqrt{2mN}}+\frac{e^{u-\mu}-1}{P_\text{RF}},\frac{e^u-1}{P_\text{RF}}\right)}{e^{-x}\times}\nonumber\\&
\,\,\,\,\,\,\,\,\,\,\,\,\,\,\,\,\,\,\,\,\,\,\,\,\,\,\,\,\,\,\,\,\,\,\,\,\,\,\,\,\,\,\,\,\,\,\,\left(\frac{1}{2}-\frac{\sqrt{mN}P_\text{RF}e^{\mu-u}(x-\frac{e^{u-\mu}-1}{P_\text{RF}})}{\sigma\sqrt{2\pi}}\right)\text{d}x
\nonumber\\&
=1-e^{-\max\left(0,\frac{-\sigma\sqrt{\pi}e^{u-\mu}}{P_\text{RF}\sqrt{2mN}}+\frac{e^{u-\mu}-1}{P_\text{RF}}\right)}\nonumber\\&+\left(\frac{1}{2}-\frac{\sqrt{mN}(e^{\mu-u}-1)}{\sigma\sqrt{2\pi}}\right) \Bigg(e^{-\max\left(0,\frac{-\sigma\sqrt{\pi}e^{u-\mu}}{P_\text{RF}\sqrt{2mN}}+\frac{e^{u-\mu}-1}{P_\text{RF}}\right)}\nonumber\\&\,\,\,\,\,\,\,\,\,\,\,\,\,\,\,\,\,\,\,\,\,\,\,\,\,\,\,\,\,\,\,\,\,\,\,\,\,\,\,\,\,\,\,\,\,\,\,\,\,\,\,\,\,\,\,\,\,\,\,\,\,\,\,\,\,-e^{-\min\left(\frac{\sigma\sqrt{\pi}e^{u-\mu}}{P_\text{RF}\sqrt{2mN}}+\frac{e^{u-\mu}-1}{P_\text{RF}},\frac{e^u-1}{P_\text{RF}}\right)}\Bigg)\nonumber\\&
-\frac{\sqrt{mN}P_\text{RF}e^{\mu-u}}{\sigma\sqrt{2\pi}}\Bigg(  \left(1+\max\left(0,\frac{-\sigma\sqrt{\pi}e^{u-\mu}}{P_\text{RF}\sqrt{2mN}}+\frac{e^{u-\mu}-1}{P_\text{RF}}\right)\right)\nonumber\\&\,\,\,\,\,\,\,\,\,\,\,\,\,\,\,\,\,\,\,\,\,\,\,\,\,\,\,\,\,\,\,\,\,\,\,\times e^{-\max\left(0,\frac{-\sigma\sqrt{\pi}e^{u-\mu}}{P_\text{RF}\sqrt{2mN}}+\frac{e^{u-\mu}-1}{P_\text{RF}}\right)}        \nonumber\\&-\left(1+\min\left(\frac{\sigma\sqrt{\pi}e^{u-\mu}}{P_\text{RF}\sqrt{2mN}}+\frac{e^{u-\mu}-1}{P_\text{RF}},\frac{e^u-1}{P_\text{RF}}\right)\right)\nonumber\\&\,\,\,\,\,\,\,\,\,\,\,\,\,\,\,\,\,\,\,\,\,\,\,\,\,\,\,\,\,\,\,\,\,\,\times e^{-\min\left(\frac{\sigma\sqrt{\pi}e^{u-\mu}}{P_\text{RF}\sqrt{2mN}}+\frac{e^{u-\mu}-1}{P_\text{RF}},\frac{e^u-1}{P_\text{RF}}\right)}                      \Bigg)=\mathcal{F}(u).
\end{align}
Here, $V(x)$ is obtained by applying the same linearization technique as in (\ref{eq:linearizationnik}) on the Gaussian $Q$ function of (\ref{eq:cdfWmapprox}) at point $x=\frac{e^{u-\mu}-1}{P_\text{RF}}$. Then, using (\ref{eq:cdfWmapproxlemma}) in (\ref{eq:eqeta1})-(\ref{eq:eqoutprob1}), one can find the throughput and outage probability, as stated in the lemma.
\end{proof}
As the second-order approximation of Lemma 1, the outage expression (15) is rephrased as
\begin{align}\label{eq:eqnewapprox12}
F_{W_m}(u)\simeq 1-e^{\frac{e^{u-\mu}-1}{P_\text{RF}}},
\end{align}
for large values of $N$. Thus, using (\ref{eq:mueq}) and $-e^{\frac{1}{x}}\text{Ei}(-\frac{1}{x})\simeq \log(x)$ for large values of $x$ in (\ref{eq:eqnewapprox12}), it is found that at high SNRs the outage probability of the joint RF-FSO link decreases with the power of RF and FSO signals exponentially, if the FSO link follows an exponential distribution.

Along with the approximation scheme of Lemma 1, Lemmas 2-5 derive upper and lower bounds of the system performance assuming that the mean and variance of the equivalent Gaussian random variable $\mathcal{Z}$ are calculated accurately.

\textbf{\emph{Lemma 2:}} The performance of the RF-FSO system is upper-estimated, i.e., the throughput is upper bounded and the outage probability is lower bounded, via the following inequality
\begin{align}\label{eq:eqlemma2}
F_{W_{m}}(u)\ge \mathcal{V}(u),
\end{align}
with $\mathcal{V}(u)$ given in (\ref{eq:eqprooflemma2}).
\begin{proof}
As mentioned before and in \cite{throughputdef,a01661837,MIMOARQkhodemun,tuninetti2011}, the throughput (resp. the outage probability) of the HARQ-based systems is a decreasing (resp. increasing) function of the probabilities $F_{W_{m}}(\frac{R}{m}),\forall m$ (resp. $F_{W_M}(\frac{R}{M})$). Thus, the throughput (resp. the outage probability) is upper bounded (resp. lower bounded) by lower bounding $F_{W_{m}}(.),\forall m.$ On the other hand, because the $Q$ function  is a decreasing function and $r(x)=\frac{\sqrt{mN}(\log(1+P_\text{RF}x)+\mu-u)}{\sigma}$ is  concave in $x$, the CDFs of the AMIs are lower bounded if $r(x)$ is replaced by its first order Taylor expansion at any point. Considering the Taylor expansion of $r(x)$ at $x=\frac{e^{u-\mu}-1}{P_\text{RF}}$, we can write
\begin{align}\label{eq:eqprooflemma2}
&F_{W_{m}}(u)\ge \nonumber\\& \int_0^{\frac{e^u-1}{P_\text{RF}}}{e^{-x}Q\left(\frac{P_\text{RF}\sqrt{mN}e^{\mu-u}}{\sigma}\left(x-\frac{e^{u-\mu}-1}{P_\text{RF}}\right)\right)\text{d}x}\nonumber\\&
\mathop  = \limits^{(h)} Q\left(\frac{\sqrt{mN}(e^{\mu-u}-1)}{\sigma}\right)\nonumber\\&-e^{-\frac{e^u-1}{P_\text{RF}}}Q\left(\frac{P_\text{RF}\sqrt{mN}e^{\mu-u}}{\sigma}\left(\frac{e^u-1}{P_\text{RF}}-\frac{e^{u-\mu}-1}{P_\text{RF}}\right)\right)\nonumber\\&
-\frac{P_\text{RF}\sqrt{mN}e^{\mu-u}}{\sigma\sqrt{2\pi}}\int_0^{\frac{e^u-1}{P_\text{RF}}}{e^{-\left(x+{\frac{P_\text{RF}^2{mN}e^{2(\mu-u)}}{2\sigma^2}}\left(x-\frac{e^{u-\mu}-1}{P_\text{RF}}\right)^2\right)}\text{d}x}\nonumber\\&
=Q\left(\frac{\sqrt{mN}(e^{\mu-u}-1)}{\sigma}\right)\nonumber\\&-e^{-\frac{e^u-1}{P_\text{RF}}}Q\left(\frac{P_\text{RF}\sqrt{mN}e^{\mu-u}}{\sigma}\left(\frac{e^u-1}{P_\text{RF}}-\frac{e^{u-\mu}-1}{P_\text{RF}}\right)\right)\nonumber\\&
+\frac{1}{2}e^{{{\frac{\sigma^2}{2P_\text{RF}^2{mN}e^{2(\mu-u)}}}}-\frac{e^{u-\mu}-1}{P_\text{RF}}}\times\nonumber\\&\Bigg(\text{erf}\left(\frac{{\frac{P_\text{RF}{mN}e^{2(\mu-u)}}{\sigma^2}}({e^{u-\mu}}-{{e^u}})-1}{{{\frac{P_\text{RF}\sqrt{2mN}e^{(\mu-u)}}{\sigma}}}}\right)
\nonumber\\&\,\,\,\,\,\,\,\,\,\,\,\,\,\,\,\,\,\,\,\,\,\,\,\,\,\,-\text{erf}\left(\frac{{\frac{P_\text{RF}{mN}e^{2(\mu-u)}}{\sigma^2}}({e^{u-\mu}-1})-1}{{{\frac{P_\text{RF}\sqrt{2mN}e^{(\mu-u)}}{\sigma}}}}\right)\Bigg)=\mathcal{V}(u).
\end{align}
Here, $(h)$ comes from  partial integration and $\frac{\text{d}Q(y(x))}{\text{d}x}=\frac{-1}{\sqrt{2\pi}}\frac{\text{d}y}{\text{d}x}e^{-\frac{y^2(x)}{2}}.$ Also, the last equality is obtained by some manipulations and the definition of the error function.
\end{proof}

\textbf{\emph{Lemma 3:}} An under-estimate of the performance of the RF-FSO system is given by
\begin{align}\label{eq:eqlemma3}
F_{W_{m}}(u)\le \mathcal{T}(u),
\end{align}
where $\mathcal{T}(u)$ is defined in (\ref{eq:eqprooflemma3}).
\begin{proof}
To derive an under-estimate of the system performance, i.e., a lower bound of the throughput and an upper bound of the outage probability, we use (\ref{eq:cdfWmapprox}) to upper bound the probability terms $F_{W_{m}}(u),\forall m,$ by
\begin{align}\label{eq:eqprooflemma3}
&F_{W_{m}}(u)\mathop  = \limits^{(i)}Q\left(\frac{\sqrt{mN}(\mu-u)}{\sigma}\right)-e^{-\frac{e^u-1}{P_\text{RF}}}Q\left(\frac{\sqrt{mN}\mu}{\sigma}\right)
\nonumber\\&-\frac{P_\text{RF}\sqrt{mN}}{\sqrt{2\pi}\sigma}\int_0^{\frac{e^u-1}{P_\text{RF}}}{\frac{e^{-x}}{1+P_\text{RF}x}e^{-\frac{{mN}\left(\log(1+P_\text{RF}x)+\mu-u\right)^2}{2\sigma^2}}\text{d}x}\nonumber\\&
\mathop  \le \limits^{(j)} Q\left(\frac{\sqrt{mN}(\mu-u)}{\sigma}\right)-e^{-\frac{e^u-1}{P_\text{RF}}}Q\left(\frac{\sqrt{mN}\mu}{\sigma}\right)
\nonumber\\&-\frac{\sqrt{mN}e^{\frac{1}{P_\text{RF}}-\frac{-mN\epsilon(\mu-u-\frac{\epsilon}{2})}{\sigma^2}}}{\sqrt{2\pi}\sigma}\int_1^{{e^u}}{\frac{e^{-\frac{t}{{P_\text{RF}}}}}{t^{1+\frac{mN\epsilon}{\sigma^2}}}\text{d}t}\nonumber\\&
=Q\left(\frac{\sqrt{mN}(\mu-u)}{\sigma}\right)-e^{-\frac{e^u-1}{P_\text{RF}}}Q\left(\frac{\sqrt{mN}\mu}{\sigma}\right)
\nonumber\\&-\frac{\sqrt{mN}e^{\frac{1}{P_\text{RF}}-\frac{-mN\epsilon(\mu-u-\frac{\epsilon}{2})}{\sigma^2}}}{\sqrt{2\pi}\sigma}\times\nonumber\\&\Bigg(E_{1+\frac{mN\epsilon}{\sigma^2}}\left(\frac{1}{P_\text{RF}}\right)-e^{-{\frac{mN\epsilon u}{\sigma^2}}}E_{1+\frac{mN\epsilon}{\sigma^2}}\left(\frac{e^u}{P_\text{RF}}\right)\Bigg)=\mathcal{T}(u).
\end{align}
In (\ref{eq:eqprooflemma3}), $(i)$ is based on partial integration. Also, $(j)$ follows from $(a-b)^2\ge \max(0,2a\epsilon-2b\epsilon-\frac{\epsilon^2}{2}),\forall a\ge b,\epsilon\ge 0,$ and variable transform $t=1+P_\text{RF}x$. Finally, the last equality is obtained by manipulations and the definition of the $n$-th order exponential integral function $E_n(x)=\int_1^{\infty}{t^{-n}e^{-tx}\text{d}t}$. Note that the bound is reasonably tight
for different values of $\epsilon\ge 0$. Then, the appropriate value of $\epsilon$ can be determined numerically such that the difference between the exact and the bounded probabilities is minimized.
\end{proof}
\textbf{\emph{Lemma 4:}} The performance of the RF-FSO system is upper-estimated via the following inequality
\begin{align}\label{eq:eqlemma4}
F_{W_{m}}(u)\ge \mathcal{R}(u),
\end{align}
with $\mathcal{R}(u)$ given in (\ref{eq:eqprooflemma4}).
\begin{proof}
Using $\log(1+x)\le x,\forall x\ge 0,$ and the same arguments as in Lemmas 2-3, the upper-estimate is found as
\begin{align}\label{eq:eqprooflemma4}
&F_{W_{m}}(u)\ge \int_0^{\frac{e^u-1}{P_\text{RF}}}{e^{-x}Q\left(\frac{\sqrt{mN}\left(P_\text{RF}x+\mu-u\right)}{\sigma}\right)\text{d}x}\nonumber\\&
=Q\left(\frac{\sqrt{mN}}{\sigma}({u-\mu})\right)-e^{-\frac{e^u-1}{P_\text{RF}}}Q\left(\frac{\sqrt{mN}}{\sigma}\left({e^u-1}+{u-\mu}\right)\right)\nonumber\\&
+\frac{1}{2}e^{{\frac{\sigma^2}{2{mN}P_\text{RF}^2}}-\frac{\mu-u}{P_\text{RF}}}\Bigg(\text{erf}\left(\frac{\frac{{mN}P_\text{RF}}{\sigma^2}({\mu-u}-{e^u+1})-1}{\frac{\sqrt{2mN}P_\text{RF}}{\sigma}}\right)\nonumber\\&\,\,\,\,\,\,\,\,\,\,\,\,\,\,\,\,\,\,\,\,\,\,\,\,\,\,\,-\text{erf}\left(\frac{\frac{{mN}P_\text{RF}}{\sigma^2}{(\mu-u)}-1}{\frac{\sqrt{2mN}P_\text{RF}}{\sigma}}\right)\Bigg)=\mathcal{R}(u),
\end{align}
where the last equality is obtained with the same procedure as in (\ref{eq:eqprooflemma2}). Note that, due to the considered bounding approach, the under-estimate is tight at low SNRs of the RF link.
\end{proof}

\textbf{\emph{Lemma 5:}} An under-estimate of the performance of the RF-FSO system is given by
\begin{align}\label{eq:eqlemma5}
F_{W_{m}}(u)\le \mathcal{S}(u),
\end{align}
where $\mathcal{S}(u)$ is defined in (\ref{eq:eqprooflemma5}).
\begin{proof}
To derive an under-estimate of the system performance,
we use (\ref{eq:cdfWmapprox}) to upper bound the probability terms $F_{W_{m}}(u),\forall m,$ by
\begin{align}\label{eq:eqprooflemma5}
&F_{W_{m}}(u)\mathop  = \limits^{(k)}Q\left(\frac{\sqrt{mN}(\mu-u)}{\sigma}\right)-e^{-\frac{e^u-1}{P_\text{RF}}}Q\left(\frac{\sqrt{mN}\mu}{\sigma}\right)
\nonumber\\&-\frac{P_\text{RF}\sqrt{mN}}{\sqrt{2\pi}\sigma}\int_0^{\frac{e^u-1}{P_\text{RF}}}{\frac{e^{-x}}{1+P_\text{RF}x}e^{-\frac{{mN}\left(\log(1+P_\text{RF}x)+\mu-u\right)^2}{2\sigma^2}}\text{d}x}\nonumber\\&
\mathop  \le \limits^{(l)} Q\left(\frac{\sqrt{mN}(\mu-u)}{\sigma}\right)-e^{-\frac{e^u-1}{P_\text{RF}}}Q\left(\frac{\sqrt{mN}\mu}{\sigma}\right)
\nonumber\\&-\frac{P_\text{RF}\sqrt{mN}}{\sqrt{2\pi}\sigma}\Bigg(\int_0^{\max(0,\frac{u-\mu}{2P_\text{RF}})}{\frac{e^{-x}e^{\frac{mN}{\sigma^2}(u-\mu)P_\text{RF}x-\frac{{mN}(u-\mu)^2}{2{\sigma^2}}}}{1+P_\text{RF}x}\text{d}x}\nonumber\\&\,\,\,\,\,\,\,\,\,\,\,\,\,\,\,\,\,\,\,\,\,\,\,\,\,\,\,\,\,\,\,\,\,\,+\int_{\max\left(0,\frac{u-\mu}{2P_\text{RF}}\right)}^{\min\left(\frac{e^u-1}{P_\text{RF}},\frac{u-\mu}{2P_\text{RF}}\right)}{\frac{e^{-x}}{1+P_\text{RF}x}\text{d}x}\Bigg)\nonumber\\&
=Q\left(\frac{\sqrt{mN}(\mu-u)}{\sigma}\right)-e^{-\frac{e^u-1}{P_\text{RF}}}Q\left(\frac{\sqrt{mN}\mu}{\sigma}\right)
\nonumber\\&-\frac{P_\text{RF}\sqrt{mN}}{\sqrt{2\pi}\sigma}\Bigg(\frac{e^{\frac{(1-\frac{mN}{\sigma^2}(u-\mu)P_\text{RF})}{P_\text{RF}}-\frac{{mN}(u-\mu)^2}{2{\sigma^2}}}}{P_\text{RF}}\times\nonumber\\&\bigg(\text{Ei}\left(-\left(1-\frac{mN}{\sigma^2}(u-\mu)P_\text{RF}\right)\left(\max\left(0,\frac{u-\mu}{2P_\text{RF}}\right)+\frac{1}{P_\text{RF}}\right)\right)\nonumber\\&-\text{Ei}\bigg(\frac{(\frac{mN}{\sigma^2}(u-\mu)P_\text{RF}-1)}{P_\text{RF}}\bigg)\bigg)\nonumber\\&
+\frac{e^{\frac{1}{P_\text{RF}}}}{P_\text{RF}}\bigg(\text{Ei}\left(-\min\left(\frac{e^u-1}{P_\text{RF}},\frac{u-\mu}{2P_\text{RF}}\right)-\frac{1}{P_\text{RF}}\right)\nonumber\\&\,\,\,\,\,\,\,\,\,\,\,\,-\text{Ei}\left(-\max\left(0,\frac{u-\mu}{2P_\text{RF}}\right)-\frac{1}{P_\text{RF}}\right)\bigg)
\Bigg)=\mathcal{S}(u).
\end{align}
In (\ref{eq:eqprooflemma5}), $(k)$ comes from partial integration. Then, $(l)$ is based on the inequality
\begin{align}\label{eq:eqlemma52}
\forall a\ge0 ,b,\, -\frac{a}{2}(\log&(1+P_\text{RF}x)-b)^2\nonumber\\&\le
\left\{\begin{matrix}
abP_\text{RF}x-\frac{ab^2}{2},\,\,\,\,\,\,\,\,\, \text{if}\,\,x<\frac{b}{2P_\text{RF}}\\
1,\,\,\,\,\,\,\,\,\,\,\,\,\,\,\,\,\,\,\,\,\,\,\,\,\,\,\,\,\,\,\,\,\,\,\,\,\,\, \text{if}\,\,x\ge \frac{b}{2P_\text{RF}}
\end{matrix}\right.
\end{align}
and the last equality is derived by the definition of the exponential integral function $\text{Ei}(x)=\int_x^\infty{\frac{e^{-t}\text{d}t}{t}}$.
\end{proof}

\subsection{Performance Analysis in the Cases with Comparable Coherence Times of the RF and FSO Links}
Up to now, the results were presented for the cases with considerably shorter coherence time of the FSO link, compared to the RF link, motivated by the results of, e.g., \cite{5342330,1142964,598416,598420}, such that the CLT provides accurate approximation for the sum of independent and identically distributed (IID) random variables. However, it is interesting to analyze the system performance in the cases with comparable coherence times of the RF and FSO links, i.e., with small values of $N$ in (\ref{eq:eqWm1}).

Here, we mainly concentrate on the Gamma-Gamma distribution of the FSO link. The same results as in \cite{5357980} can be applied to derive the CDFs $F_{W_{m}}$ with, e.g., the exponential distribution of FSO link and small $N.$


Using the Minkowski inequality \cite[Theorem 7.8.8]{minkowskibook}
\begin{align}\label{eq:eqminko}
\left(1+\left(\prod_{i=1}^n{x_i}\right)^\frac{1}{n}\right)^n\le \prod_{i=1}^{n}{\left(1+x_i\right)},
\end{align}
we have
\begin{align}\label{eq:eqminkowseq1}
&\Pr\left(\frac{1}{mN}\sum_{j=1}^m\sum_{i=1}^N{\log(1+P_\text{FSO}G_{\text{FSO},1+(j-1)i})}\le x\right)\nonumber\\&= \Pr\left(\prod_{j=1}^m\prod_{i=1}^N(1+P_\text{FSO}G_{\text{FSO},1+(j-1)i})\le e^{mNx}\right)\nonumber\\&\le \Pr\left(1+P_\text{FSO}\left(\prod_{j=1}^m\prod_{i=1}^N{G_{\text{FSO},1+(j-1)i}}\right)^{\frac{1}{mN}}\le e^{x}\right)\nonumber\\&=F_\mathcal{Q}\left(\left(\frac{e^x-1}{P_\text{FSO}}\right)^{mN}\right),
\end{align}
where using the results of \cite[Lemma 3]{6168189} and for the Gamma-Gamma distribution of the variables $G_{\text{FSO},1+(j-1)i}$, $\mathcal{Q}=\prod_{j=1}^m\prod_{i=1}^N{G_{\text{FSO},1+(j-1)i}}$ follows the CDF
\begin{align}\label{eq:eqresminkows}
F_\mathcal{Q}(x)&=\frac{1}{\Gamma^{mN}(a)\Gamma^{mN}(b)}\times\nonumber\\&\mathcal{G}_{1,2mN+1}^{2mN,1}\Bigg(({ab})^{mN}x\Bigg|_{\underbrace{a,a,\ldots,a}_{mN \text{ times}},\,\underbrace{b,b,\ldots,b}_{mN \text{ times}},0}^{\,\,\,\,\,\,\,\,1}\Bigg),
\end{align}
with $\mathcal{G}(.)$ denoting the Meijer G-function.

In this way, from (\ref{eq:eqWm1}) and (\ref{eq:eqresminkows}), the probabilities $\Pr(W_m\le u),\forall m,$ are upper-bounded by
\vspace{-3mm}
\begin{align}\label{eq:equpperboundG}
&\Pr(W_m\le u)\le \nonumber\\& \frac{1}{\Gamma^{mN}(a)\Gamma^{mN}(b)}\int_0^{\frac{e^u-1}{P_\text{RF}}}{e^{-x}\mathcal{G}_{1,2mN+1}^{2mN,1}\Bigg(}\nonumber\\&
\bigg(\frac{ab(e^{u-\log(1+P_\text{RF}x)}-1)}{ P_\text{FSO}}\bigg)^{mN}\Bigg|_{\underbrace{a,a,\ldots,a}_{mN \text{ times}},\,\underbrace{b,b,\ldots,b}_{mN \text{ times}},0}^{\,\,\,\,\,\,\,\,1}\Bigg)\text{d}x
\end{align}
which
 can be calculated numerically.

On the other hand, using the Jensen's inequality and the concavity of the logarithm function, we can write
\begin{align}\label{eq:eqjensen11}
\frac{1}{n}\sum_{i=1}^n{\log\left(1+x_i\right)}\le \log\left(1+\frac{1}{n}\sum_{i=1}^n{x_i}\right),
\end{align}
which leads to
\begin{align}\label{eq:eqjenseneq1}
&\Pr\left(\frac{1}{mN}\sum_{j=1}^m\sum_{i=1}^N{\log(1+P_\text{FSO}G_{\text{FSO},1+(j-1)i})}\le x\right)\nonumber\\&\ge \Pr\left(\mathcal{B}\le \frac{mN}{P_\text{FSO}}(e^x-1)\right)=F_\mathcal{B}\left(\frac{mN}{P_\text{FSO}}(e^x-1)\right).
\end{align}
Here, $\mathcal{B}\doteq\sum_{j=1}^m\sum_{i=1}^N{G_{\text{FSO},1+(j-1)i}}$ follows the PDF \cite[Lemma 1]{6168189}\footnote{Equation (\ref{eq:eqmollalemma1}) gives an approximation of the sum of IID Gamma-Gamma variables. However, as shown in, e.g., \cite{6168189} the approximation is extremely tight for all ranges of SNRs. Therefore, we consider it as an equality.}
\begin{align}\label{eq:eqmollalemma1}
&f_\mathcal{B}(x)=\frac{2}{\Gamma(\sigma_{mN})\Gamma(\varsigma_{mN})x}\times\nonumber\\&\left(\frac{\sigma_{mN}\varsigma_{mN}}{mN}x\right)^{\frac{\sigma_{mN}+\varsigma_{mN}}{2}}K_{\sigma_{mN}-\varsigma_{mN}}\left(2\sqrt{\frac{\sigma_{mN}\varsigma_{mN}}{mN}x}\right),\nonumber\\&
\sigma_{mN}=mN\upsilon+\varrho_{mN},\nonumber\\&
\varsigma_{mN}=mN\tau,\nonumber\\&
\tau=\min\{a,b\},
\upsilon=\max\{a,b\},
\end{align}
with $\varrho_{mN}$ being an appropriately chosen adjustment parameter.

From (\ref{eq:eqWm1}) and (\ref{eq:eqmollalemma1}), the probabilities $\Pr(W_m\le u),\forall m,$ are lower-bounded by
\begin{align}\label{eq:eqlowermolla}
&\Pr(W_m\le u)\ge \nonumber\\&\int_0^{u}{\frac{mN}{P_\text{FSO}}e^x f_\mathcal{B}\left(\frac{mN}{P_\text{FSO}}(e^x-1)\right)\Pr\left(G_\text{RF}\le\frac{e^{u-x}-1}{P_\text{RF}}\right)\text{d}x}\nonumber\\&
=\frac{2}{\Gamma(\sigma_{mN})\Gamma(\varsigma_{mN})}\int_0^u{\frac{e^x}{e^x-1}\left(\frac{\sigma_{mN}\varsigma_{mN}}{P_\text{FSO}}(e^x-1)\right)^{\frac{\sigma_{mN}+\varsigma_{mN}}{2}} }\nonumber\\&\times K_{\sigma_{mN}-\varsigma_{mN}}\left(2\sqrt{\frac{\sigma_{mN}\varsigma_{mN}}{P_\text{FSO}}(e^x-1)}\right)\left(1-e^{-\frac{e^{u-x}-1}{P_\text{RF}}}\right)\text{d}x,
\end{align}
if the channel realizations of the FSO link follow Gamma-Gamma distribution. Particularly, using the tight high-SNR approximation of (\ref{eq:eqmollalemma1}) as \cite[Lemma 2]{6168189}
\begin{align}\label{eq:eqmollasag}
f_\mathcal{B}(x)\sim \left(\frac{\Gamma(\upsilon-\tau)({\upsilon\tau})^\tau}{\Gamma(\upsilon)}\right)^{mN}\frac{x^{mN\tau-1}}{\Gamma(mN\tau)},
\end{align}
(\ref{eq:eqlowermolla}) is rephrased as
\begin{align}\label{eq:eqxxxx}
\Pr(W_m\le u)\ge & \left(\frac{\Gamma(\upsilon-\tau)({\upsilon\tau})^\tau}{\Gamma(\upsilon)}\right)^{mN}\left(\frac{mN}{P_\text{FSO}}\right)^{mN\tau}\times\nonumber\\&\int_0^u{e^x\frac{(e^x-1)^{mN\tau-1}}{\Gamma(mN\tau)}\left(1-e^{-\frac{e^{u-x}-1}{P_\text{RF}}}\right)\text{d}x},
\end{align}
at high SNRs.


In Section IV, we validate the accuracy of the bounds/approximations proposed in (\ref{eq:mueq})-(\ref{eq:eqsigmagamma1}) by comparing them with the corresponding values obtained via simulations. Also, note that the results of (\ref{eq:equpperboundG}), (\ref{eq:eqlowermolla}) and (\ref{eq:eqxxxx}) are mathematically applicable for every value of $N$. However, for, say $N\ge 6,$ the implementation of, e.g., the Meijer G-function in MATLAB is very time-consuming and the tightness of the approximations (\ref{eq:equpperboundG}), (\ref{eq:eqlowermolla}) and (\ref{eq:eqxxxx}) decreases for large values of $N$. As a result, (\ref{eq:equpperboundG}), (\ref{eq:eqlowermolla}) and (\ref{eq:eqxxxx}) are useful for the performance analysis in the cases with small $N$'s, while the CLT-based approach of Section III.A provides accurate performance evaluation as $N$ increases.
\subsection{On the Effect of Power Allocation}

In all figures of Section IV, except Fig. 11 which evaluates the effect of power allocation, we consider uniform power allocation between the RF and the FSO links, i.e., $P_\text{RF}=P_\text{FSO}=\frac{P}{2}.$ Indeed, the system performance is improved by adaptive power allocation based on the links long-term channel conditions. For this reason, Lemmas 6-7 derive power allocation at high and low SNRs, respectively. The results of the lemmas are of interest because they hold for different channel conditions/performance metrics.

\textbf{\emph{Lemma 6:}} At high SNRs, the optimal, in terms of throughput/outage probability, power allocation between the RF and the FSO links converges to uniform power allocation, i.e. $P_\text{RF}=P_\text{FSO}$, independently of the links channel conditions.

\begin{proof}
Using $\log(1+x)=\log(x)$ at high SNRs, the AMIs (\ref{eq:eqWm1}) are rephrased as
\begin{align}\label{eq:lemmaeqWm61}
W_m&=\log(P_\text{RF}G_\text{RF})+\frac{1}{m}\sum_{j=1}^{m}{\left(\frac{1}{N}\sum_{i=1}^N{\log(P_\text{FSO}G_{\text{FSO},1+(j-1)i})}\right)}
\nonumber\\&=\log(P_\text{RF}P_\text{FSO})+ \log\left(G_\text{RF}\sqrt[mN]{\prod_{j=1}^m\prod_{i=1}^N{G_{\text{FSO},1+(j-1)i}}}\right),\forall m.
\end{align}
Hence, the throughput (\ref{eq:eqeta1}) and the outage probability (\ref{eq:eqoutprob1}) are monotonic functions of terms $F_{W_m}(\frac{R}{m})=\Pr(\log(P_\text{RF}P_\text{FSO})+ \log\left(G_\text{RF}\sqrt[mN]{\prod_{j=1}^m\prod_{i=1}^N{G_{\text{FSO},1+(j-1)i}}}\right)\le \frac{R}{m}),\forall m,$ at high SNRs, and with a sum power constraint $P_\text{RF}+P_\text{FSO}=P$, the optimal, in terms of throughput/outage probability, power allocation rule of the RF-FSO system is given by
\begin{align}\label{eq:problemformulationopt}
\left\{\begin{matrix}
\mathop {\min }\limits_{P_\text{RF}, P_\text{FSO}} \,\,\,\,\,\,\,\,\,  {\log(P_\text{RF}P_\text{FSO})}\,\,\,\,\,\,\,\,\,\,\,\,\,\,\\
\text{subject}\,\text{to}\,\,\,\,  P_\text{RF}+P_\text{FSO}=P,
\end{matrix}\right.\mathop  \Rightarrow \limits^{(m)} P_\text{RF}=P_\text{FSO}=\frac{P}{2}.
\end{align}
Here, $(m)$ is obtained by manipulations and the fact that the logarithm $\log(P_\text{RF}P_\text{FSO})$ is an increasing function of the product $P_\text{RF}P_\text{FSO}$. Also, in (\ref{eq:problemformulationopt}) we have used the fact that all subparts of the throughput and outage probability functions in (\ref{eq:eqeta1}) and (\ref{eq:eqoutprob1}), i.e., $F_{W_m}$'s, are positive and decreasing functions of $\log(P_\text{RF}P_\text{FSO})$ (and there is no other terms related to $P_\text{RF}$, $P_\text{FSO}$ in the $F_{W_m}$'s). As a result, the whole throughput/outage probability function is a monotonic function of $\log(P_\text{RF}P_\text{FSO})$. Finally, note that the conclusion is independent of the RF and the FSO links channel PDFs.
\end{proof}

\textbf{\emph{Lemma 7:}} Consider the low SNR regime and considerably shorter coherence time of the FSO link compared to the one in the RF link. Then, depending on the parameter settings, the minimum low-SNR outage probability is achieved by using only the RF or the FSO link.

\begin{proof}
Using $\log(1+x)= x$ for small $x$'s, the AMI (\ref{eq:eqWm1}) is approximated as
\begin{align}\label{eq:eqlemmalowsnr7}
W_m=P_\text{RF}G_\text{RF}+\frac{1}{mN}\sum_{j=1}^{m}\sum_{i=1}^N{P_\text{FSO}G_{\text{FSO},1+(j-1)i}},
\end{align}
which, from (\ref{eq:eqoutprob1}), leads to outage probability
\begin{align}\label{eq:eqoutagelowSNRlemma7}
&\Pr(\text{Outage})\nonumber\\&=\Pr\left(P_\text{RF}G_\text{RF}+\frac{1}{MN}\sum_{j=1}^{M}\sum_{i=1}^N{P_\text{FSO}G_{\text{FSO},1+(j-1)i}}\le \frac{R}{M}\right)\nonumber\\&\mathop  \simeq \limits^{(n)} 1-e^{-\frac{\frac{R}{M}-P_\text{FSO}\mu_\text{FSO}}{P_\text{RF}}}.
\end{align}
Here, $(n)$ is based on the fact that with considerably different coherence times of the RF and FSO links, i.e., large values of $N$, we have $\frac{1}{MN}\sum_{j=1}^{M}\sum_{i=1}^N{P_\text{FSO}G_{\text{FSO},1+(j-1)i}}\simeq  \mu_\text{FSO},\forall M\ge 1,$ where $\mu_\text{FSO}\doteq E\{G_\text{FSO}\}$ is the mean of the channel gain in the FSO link. From (\ref{eq:eqoutagelowSNRlemma7}), the low-SNR outage-optimized power allocation problem can be formulated as
\begin{align}\label{eq:ASSTCproblems}
\left\{ \begin{array}{l}
 \mathop {\min }\limits_{ P_\text{FSO},P_\text{RF}} \,\, \Pr(\text{Outage}), \\
 \text{s.t.}\,\,\,\,  P_\text{FSO}+P_\text{RF}=P \\
 \end{array} \right.\equiv\left\{ \begin{array}{l}
 \mathop {\min }\limits_{  P_\text{FSO},P_\text{RF}} \,\, \frac{\frac{R}{M}-P_\text{FSO}\mu_\text{FSO}}{P_\text{RF}},\,\, \text{(i)} \\
 \text{s.t.}\,\,\,\,  P_\text{FSO}+P_\text{RF}=P. \,\, \text{(ii)}\\
 \end{array} \right.
\end{align}
In this way, setting $P_\text{RF}=P-P_\text{FSO}$ in (\ref{eq:ASSTCproblems}.i), the outage-optimized power allocation is given by optimizing the function $y\left(P_\text{FSO}\right)=\frac{\frac{R}{M}-P_\text{FSO}\mu_\text{FSO}}{P-P_\text{FSO}}$. Then, because $y\left(P_\text{FSO}\right)$ is a decreasing (resp. an increasing) function of $P_\text{FSO}$ for $R\ge MP\mu_\text{FSO}$ (resp. $R< MP\mu_\text{FSO}$), the outage-optimized set of powers is found as $(P_\text{RF},P_\text{FSO})=(P,0)$ for $R\ge MP\mu_\text{FSO}$ and $(P_\text{RF},P_\text{FSO})=(0,P)$ for $R< MP\mu_\text{FSO}$. That is, at low SNRs, depending on the parameter settings, the minimum outage probability is achieved by using only one of the RF or the FSO links, as stated in the lemma.
\end{proof}
Intuitively, Lemma 7 indicates that at low SNRs the power resources are limited and the RF-FSO system is conservative. In this case, because the AMI of FSO link converges to its ergodic capacity for considerably different coherence times of the links, if the ergodic capacity of the FSO link supports the code rate, all available power is assigned to the FSO link. On the other hand, if the FSO link experiences very poor channel conditions and the SNR is low, the minimum outage probability is achieved by allocating all available power to the RF link. At high SNRs, however, the RF-FSO setup can benefit from the network diversity, and the optimal power allocation tends towards uniform power allocation, as shown in Lemma 6.

Note that in Lemmas 6-7 we ignored the individual power constraints of the RF and FSO links and concentrated on the cases with a sum power constraint on the joint RF-FSO system. This scenario is of interest in the green communication concept, where the goal is to minimize the total power required for data transmission \cite{7088654}, and also for electricity-bill minimization. In practice, however, one should also consider the individual properties/constraints of the RF and FSO links such as the eye safety constraints of the FSO link and the nonlinearity/saturation conditions of the optical and RF devices.

Finally, it is interesting to note that 1) as previously proved in  \cite[Section V.B]{MIMOARQkhodemun}, the same performance is achieved by the INR and repetition time diversity (RTD) HARQ systems  at low SNRs. Thus, although the paper concentrates on the INR HARQ, the same conclusions hold for the RTD-based HARQ setups, as long as the SNR is low. 2) Throughout the paper, we considered sufficiently long codes such that the achievable rate of a link is given by $\log(1+x)$ with $x$ standing for the link SNR. Then, as shown in \cite{letterzorzikhodemun}, the performance of the HARQ codes with asymptotically long codes is very close to the ones with finite block-length. Therefore, although the results of the paper are obtained for the cases with long codewords, similar results are expected in the cases with sub-codewords of moderate length. 3) The paper concentrates on the RF-FSO based systems. However, the same system model as in Figs. 1-2 holds in various coordinated data transmission schemes, for which the analytical results of Section III is useful. Particularly, the performance analysis of RF-based coordinated nodes operating at different carrier frequencies, and the combination of radio over FSO (RoFSO) \cite{5464339} links with RF/FSO links are interesting research topics, for which our techniques can be helpful. Finally, 4) with some manipulations, we can use the derived PDFs of the AMIs and the same procedure as in \cite[Section IV.C]{7192727} to analyze the bit error rate of the RF-FSO system.

\section{Simulation Results}
Throughout the paper, we presented different approximation/bounding techniques. The verification of these results is demonstrated in Figs. 3-7 and, as seen in the sequel, the analytical results follow the simulations with high accuracy. Then, to avoid too much information in each figure, Figs. 8-11 report only the simulation results. Note that in all simulations we have double-checked the results with the ones obtained analytically. Moreover, in all figures, except for Fig. 11, we consider uniform power allocation between the RF and the FSO links, i.e., $P_\text{RF}=P_\text{FSO}=\frac{P}{2}.$ Hence, the sum power is $P$ (in dB, $10\log_{10} P$) which, because the noise variances are set to 1, is referred to the SNR as well. The effect of power allocation on the system performance is studied in Fig. 11. In Figs. 3-9, 11, we assume the FSO link to follow the exponential distribution $f_{G_\text{FSO}}(x)=\lambda_\text{FSO} e^{-\lambda_\text{FSO} x}$ with $\lambda_\text{FSO}=1,$ the log-normal distribution $f_{G_\text{FSO}}(x)=\frac{1}{\sqrt{2\pi}{\delta x}}e^{-\frac{(\log(x)-\varpi)^2}{2\delta^2}}$ with $\delta=1$ and $\varpi=0$ or the Gamma-Gamma distribution $f_{G_\text{FSO}}(x)=\frac{2(ab)^{\frac{a+b}{2}}}{\Gamma(a)\Gamma(b)}x^{\frac{a+b}{2}-1}K_{a-b}\left(2\sqrt{abx}\right),$ $a=4.3939, b=2.5636$ which corresponds to Rytov variance 1 \cite{6168189}. In Fig. 10, the RF and the FSO links are supposed to experience exponential distributions $f_{G_\text{RF}}(x)=\lambda_\text{RF} e^{-\lambda_\text{RF} x}$ and $f_{G_\text{FSO}}(x)=\lambda_\text{FSO} e^{-\lambda_\text{FSO} x}$, where $\lambda_\text{RF}$ and $\lambda_\text{FSO}$ follow normalized log-normal distributions.

\begin{figure}
\centering
  \includegraphics[width=0.99\columnwidth]{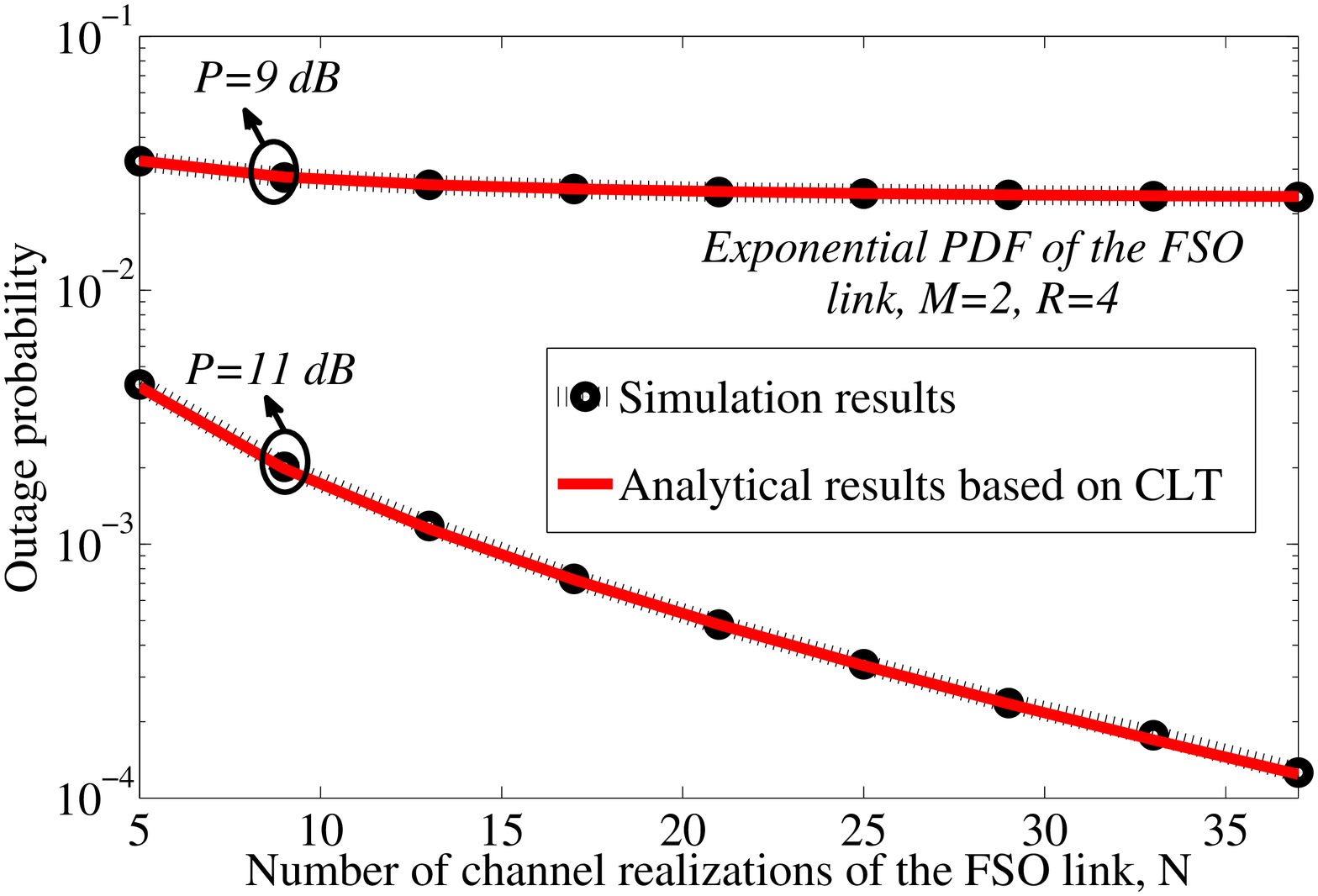}\\ \vspace{-1mm}
\caption{On the accuracy of the CLT-based approximation (exponential distribution of the FSO link, $R=4,$ and $M=2$). For every given power, the simulation results and the analytical results based on the CLT-based approximation scheme are superimposed. } \vspace{-1mm}\label{figure111}
\end{figure}

The simulation results are presented in different parts as follows.

\emph{On the effect of different coherence times:} In Fig. 3, we verify the accuracy of the CLT-based approximation and investigate the outage probability for different number of channel realizations in the FSO link. Particularly, setting $M=2$ and $R=4$ npcu, the figure compares the outage probability obtained through simulations and the result calculated from (\ref{eq:cdfWmapprox}) when $\mu$ and $\sigma$ are given by (\ref{eq:mueq}) and (\ref{eq:sigmaeq}). As demonstrated, the CLT-based approximation accurately mimics the simulations, and the difference between the analytical and simulation results is negligible even for few number of channel realizations $N$, such that the curves are superimposed (The same results hold for the other PDFs, although not demonstrated). Moreover, the outage probability decreases with increasing the number of channel realizations in the FSO link $N,$ particularly when the SNR increases. This is intuitively because more time diversity is exploited by the HARQ when the channel changes during the data transmission.

\begin{figure*}
\centering
  \includegraphics[width=1.8\columnwidth]{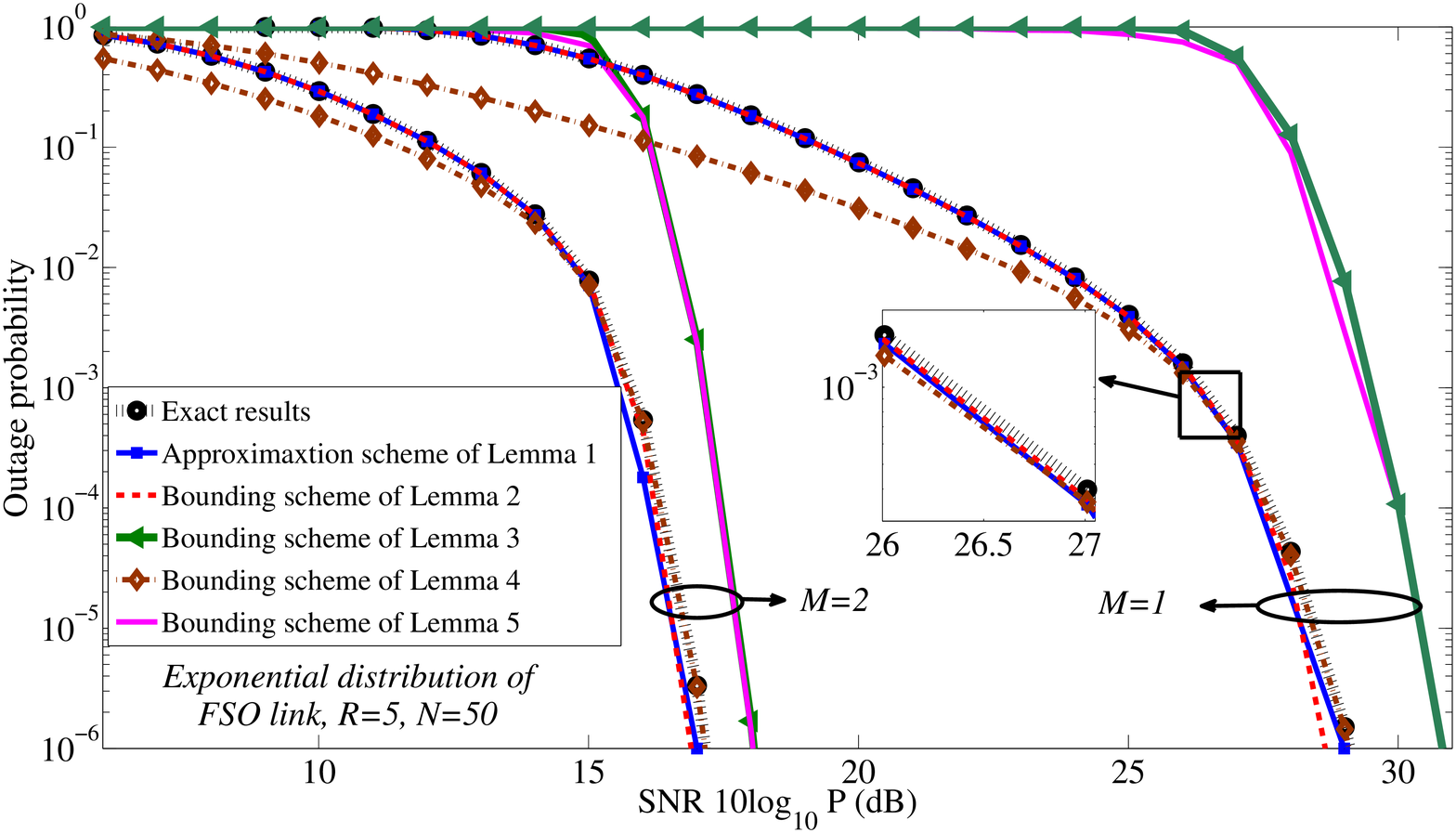}\\ \vspace{-1mm}
\caption{Comparison between the numerical and approximation results of Lemmas 1-5 in the cases with short coherence time of the FSO link (exponential distribution of the FSO link, $R=5,$ and $N=50$).} \vspace{-1mm}\label{figure111}
\end{figure*}
\begin{figure*}
\centering
  \includegraphics[width=1.8\columnwidth]{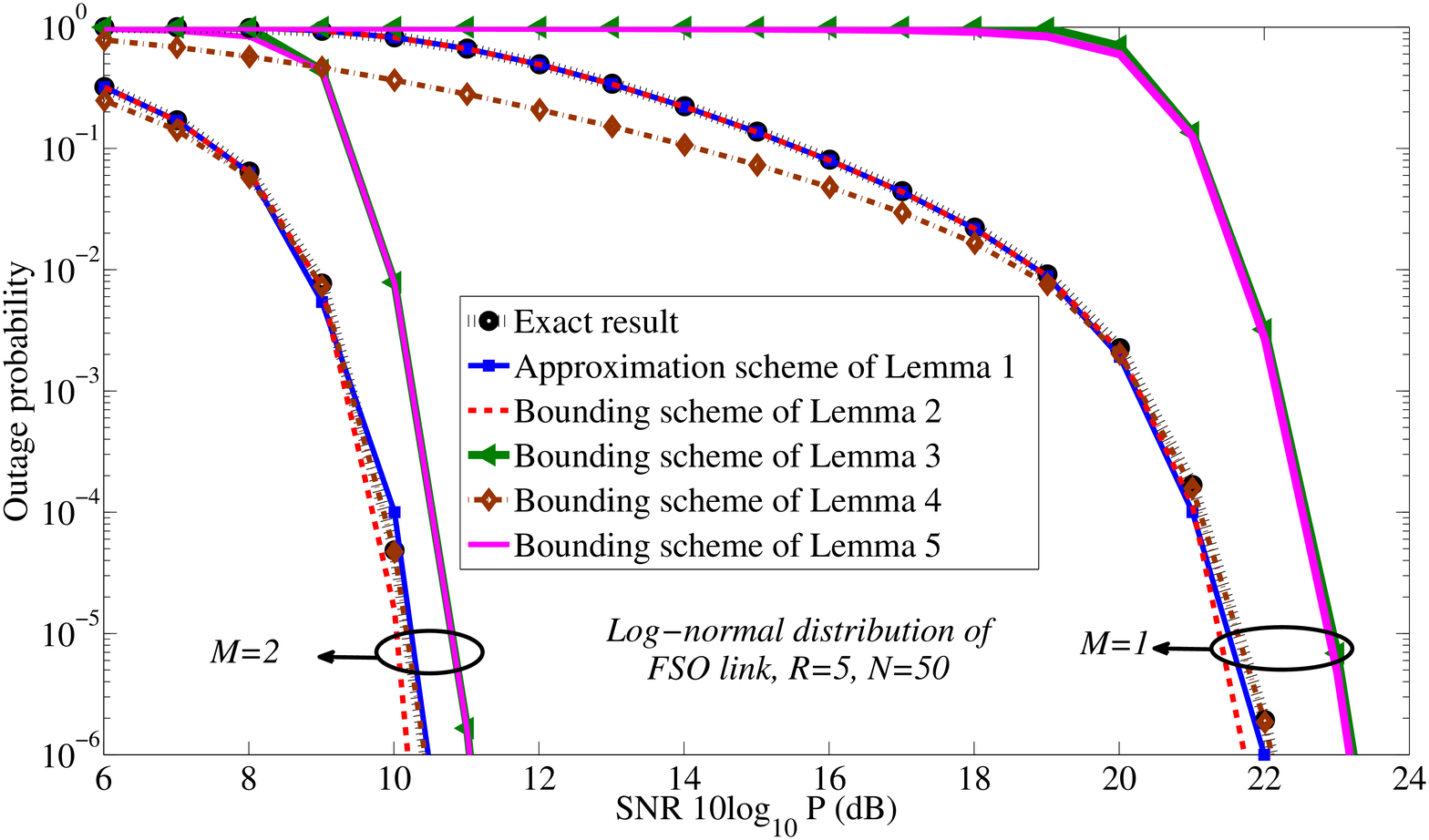}\\ \vspace{-1mm}
\caption{Comparison between the numerical and approximation results of Lemmas 1-5 in the cases with short coherence time of the FSO link (log-normal distribution of the FSO link, $R=5,$ and $N=50$).} \vspace{-1mm}\label{figure111}
\end{figure*}

\begin{figure*}
\centering
  \includegraphics[width=1.8\columnwidth]{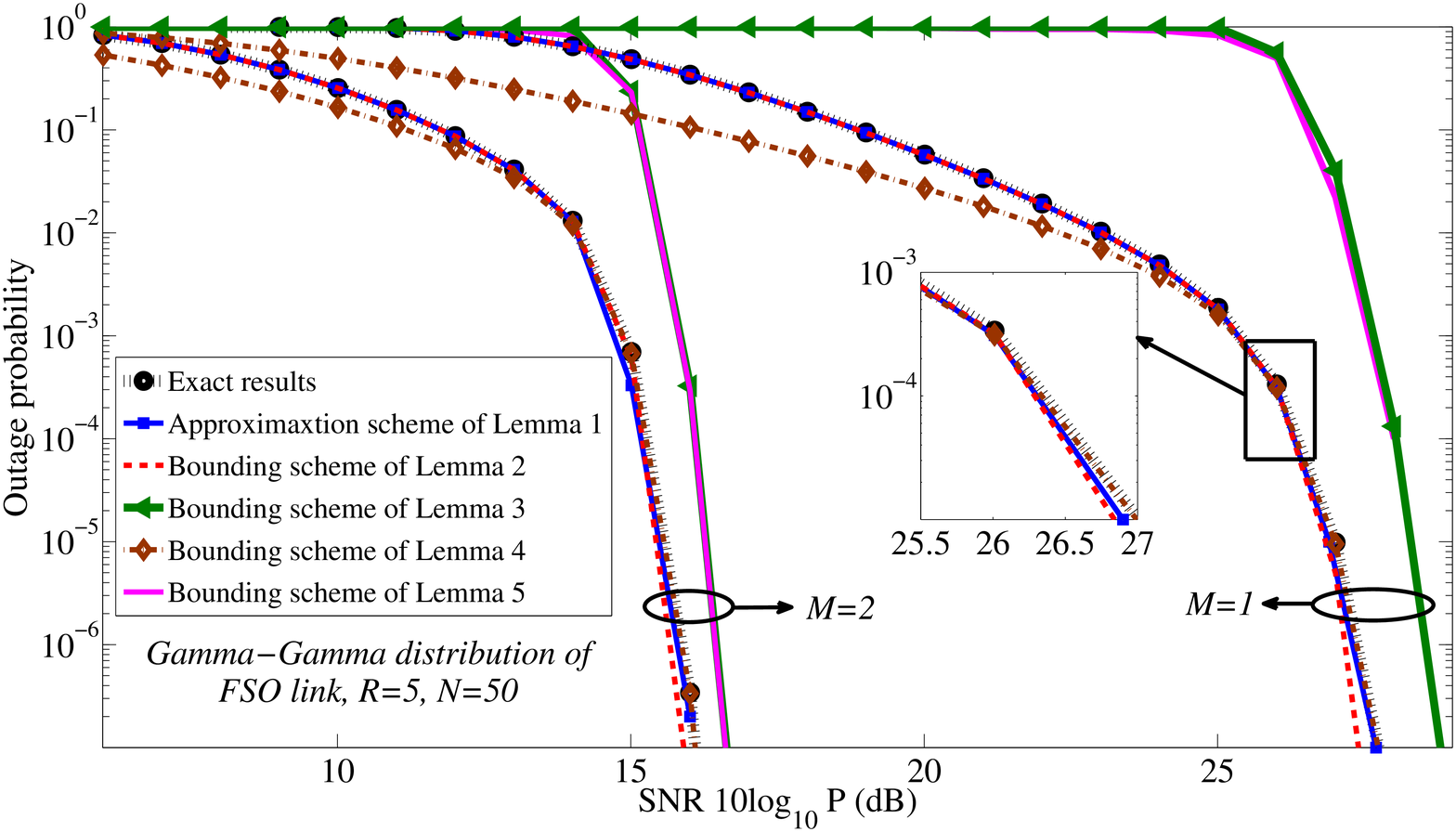}\\ \vspace{-1mm}
\caption{Comparison between the numerical and approximation results of Lemmas 1-5 in the cases with short coherence time of the FSO link (Gamma-Gamma distribution of the FSO link, $R=5,$ and $N=50$).} \vspace{-1mm}\label{figure111}
\end{figure*}

\emph{On the bounding/approximation approaches of Lemmas 1-5:} Setting $R=5$ npcu and $N=50,$ Figs. 4-6 verify the tightness of the approximation/bounding schemes of Lemmas 1-5 for the exponential, log-normal and Gamma-Gamma distributions of the FSO link, respectively. As it is observed, the analytical results of Lemmas 1, 2 and 4 mimic the exact results with very high accuracy. Also, Lemmas 3 and 5 properly upper-bound the outage probability and the tightness increases with the SNR and/or number of retransmission rounds.
In this way, according to Figs. 3-6, the CLT-based approximation approaches of Lemmas 1-5 provide effective tools for the analytical investigation of the RF-FSO systems, if the links experience different coherence times.

\emph{Performance analysis in the cases with comparable coherence times of the RF and FSO links:} Considering the Gamma-Gamma distribution of the FSO link, Fig. 7 demonstrates the outage probability of the RF-FSO system in the cases with $N=1$ instantaneous channel realization of the FSO link during retransmissions. As demonstrated, the bounds of (\ref{eq:equpperboundG}), (\ref{eq:eqlowermolla}) match the exact values derived via numerical analysis of $\Pr(W_m\le \frac{R}{m})$ exactly in the cases with $M=1$. Also, the bounding/approximation methods of (\ref{eq:equpperboundG}), (\ref{eq:eqlowermolla}) and (\ref{eq:eqxxxx}) mimic the numerical results with high accuracy in the cases with a maximum of $M=2$ retransmissions. Thus, the results of Section III.B can be efficiently used to analyze the RF-FSO systems in the cases with small values of $N$. Also, comparing Figs. 6 and 7, it is found that the relative performance gain of the HARQ, compared to open-loop communication ($M=1$), increases as the difference between the coherence times of the links increases, which is because more time diversity is exploited by the HARQ as $N$ increases in (\ref{eq:eqWm1}).

\emph{On the effect of HARQ retransmissions:} Shown in Fig. 8 are the outage probability and the throughput of the RF-FSO system for different maximum number of HARQ retransmission rounds $M$. Here, the results are presented for the exponential distribution of the FSO link, while the same trend is observed for the Gamma-Gamma and log-normal distributions of the FSO link as well. As demonstrated in the figures, the implementation of HARQ leads to significant outage probability reduction at moderate/high SNRs. On the other hand, the HARQ is more useful, in terms of throughput, at low/moderate SNRs. However, at high SNRs and with given rates, the effect of HARQ on the throughput becomes negligible, because the data is decoded successfully in the first retransmission(s) with high probability (Fig. 8b). Finally, for different distributions of the FSO link, the throughput increases with the maximum number of retransmissions $M$, and the largest relative throughput/outage probability improvement is observed when going from open-loop communication ($M=1$) to the cases with a maximum of $M=2$ retransmissions.

\emph{On the effect of initial code rate:} Figure 9 shows the throughput versus the initial code rate $R$. Here, the results of the figure are presented for the log-normal distribution of the FSO link, $N=100$ and different SNRs. As observed, for small values of $R$, the throughput increases with the rate (almost) linearly, because with high probability the data is correctly decoded in the first round. On the other hand, the outage probability increases and the throughput goes to zero for large values of $R$. Moreover, the figure indicates that the HARQ is useful, in terms of throughput, for large values of initial code rate and the relative performance gain of the HARQ increases with the SNR. Finally, depending on the SNR and the channel PDFs, there may be a number of local optimum, in terms of throughput, for the initial rate.

\emph{Comparison between the performance of the RF, the FSO and the RF-FSO based systems:} In Fig. 10, we compare the outage probability in the systems using only the RF link, only the FSO link and the joint RF-FSO transmission setup. Here, the results are obtained for the exponential PDFs of the RF and FSO links, i.e., $f_{G_\text{RF}}(x)=\lambda_\text{RF} e^{-\lambda_\text{RF} x}$ and $f_{G_\text{FSO}}(x)=\lambda_\text{FSO} e^{-\lambda_\text{FSO} x}$, where $\lambda_\text{RF}$ and $\lambda_\text{FSO}$ follow normalized log-normal distributions. Also, to have a fair comparison, the transmission powers are set to $(P_\text{RF}=P,P_\text{FSO}=0)$, $(P_\text{RF}=0,P_\text{FSO}=P)$ and $(P_\text{RF}=\frac{P}{2},P_\text{FSO}=\frac{P}{2})$ in the cases with only RF, only FSO and RF-FSO system, respectively, such that the sum power remains the same in different cases.

As demonstrated, the RF-FSO link leads to substantially less outage probability, compared to the cases with only the RF or the FSO link. For instance, with the initial rate $R=5$ npcu, $M=2$ retransmissions, and the outage probability $10^{-2}$, the joint RF-FSO based data transmission reduces the required power by $16$ and $4$ dB, compared to the cases with only the RF or the FSO link, respectively. Intuitively, this is because with the joint RF-FSO setup the diversity increases and the RF (resp. the FSO) link compensates the effect of the FSO (resp. RF) link, if it experiences poor channel conditions. Also, the effect of the joint transmission increases with the number of retransmissions/SNRs (Fig. 10). Finally, with the parameter settings of Fig. 10, the HARQ-based FSO link with $M=2$ leads to lower outage probability compared to the open-loop RF-FSO system with no HARQ feedback ($M=1$). Thus, selecting the best approach is not easy since the decision depends on several parameters such as complexity, commercial issues and the considered quality-of-service requirements.

\begin{figure}
\centering
  \includegraphics[width=0.99\columnwidth]{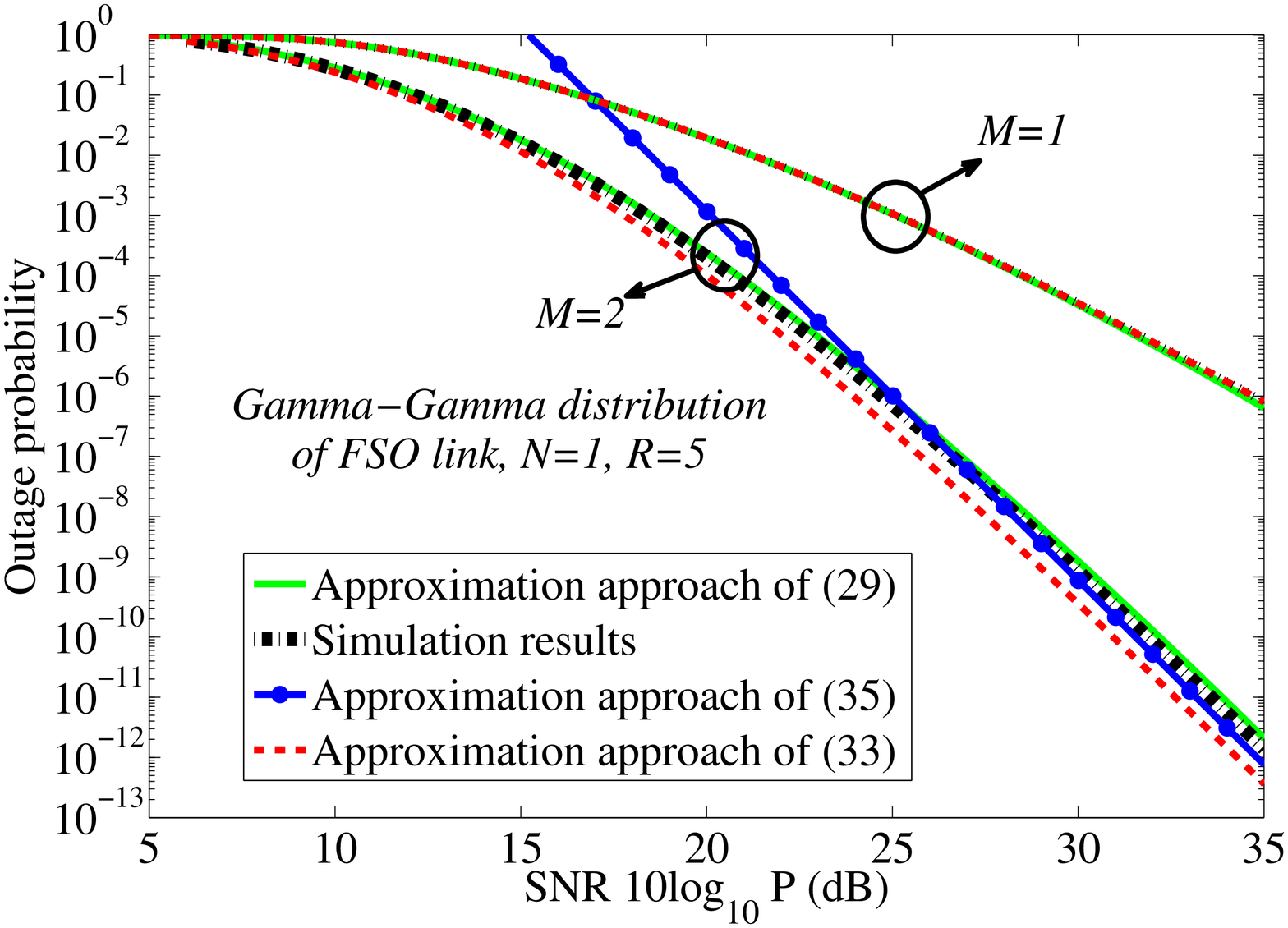}\\ \vspace{-1mm}
\caption{Comparison between the numerical and approximation results in the cases with comparable coherence times of the RF and FSO links (Gamma-Gamma distribution of the FSO link, $R=5,$ and $N=1$).} \vspace{-1mm}\label{figure111}
\end{figure}

\begin{figure}
\centering
  \includegraphics[width=0.99\columnwidth]{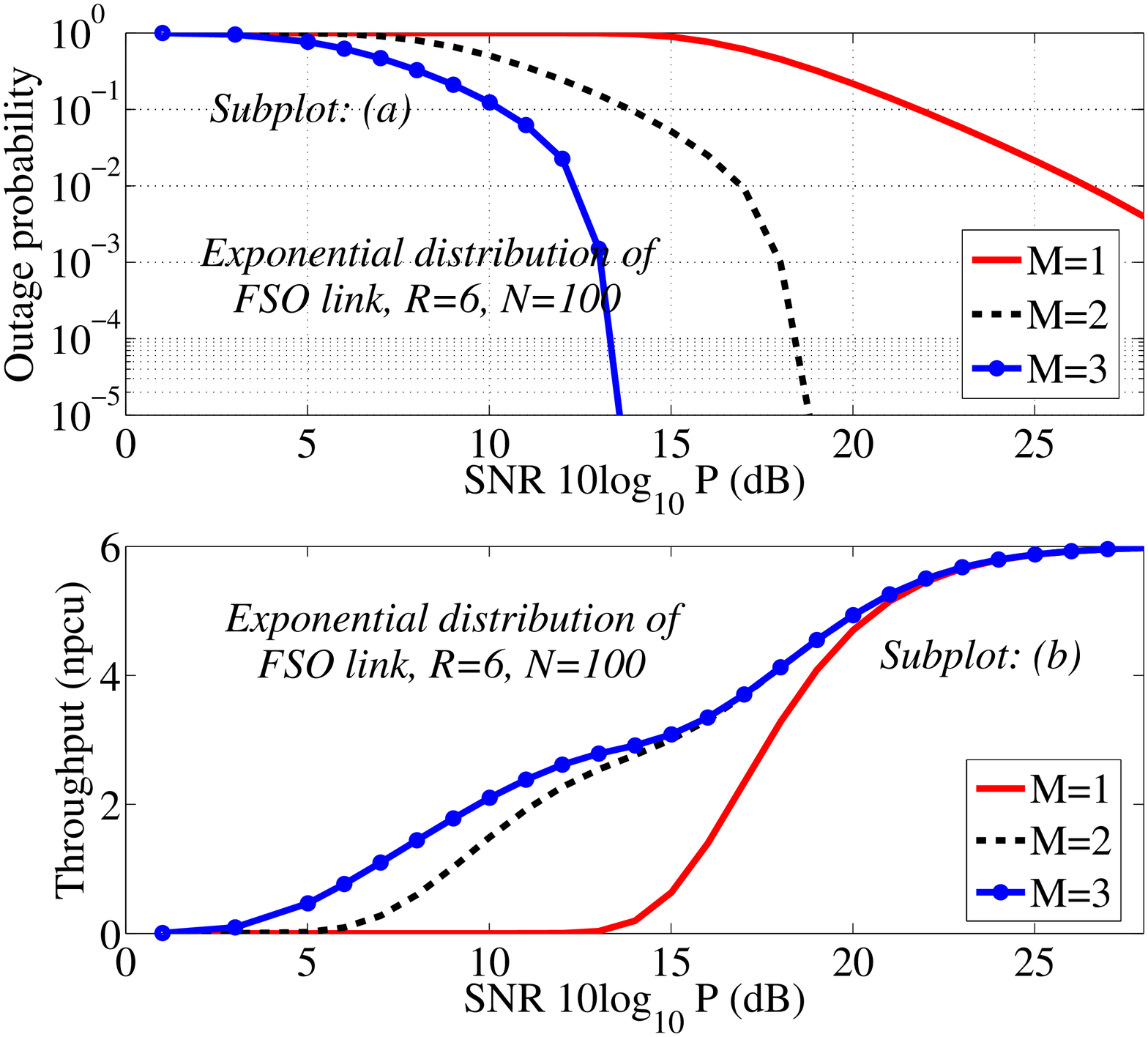}\\ \vspace{-1mm}
\caption{The effect of the number of HARQ retransmissions on the (a): outage probability, (b): throughput. Exponential PDF of FSO link, $f_{G_\text{FSO}}(x)=\lambda e^{-\lambda x},\lambda=1,$ $R=6$ npcu, and $N=100.$} \vspace{-1mm}\label{figure111}
\end{figure}

\begin{figure}
\centering
  \includegraphics[width=0.99\columnwidth]{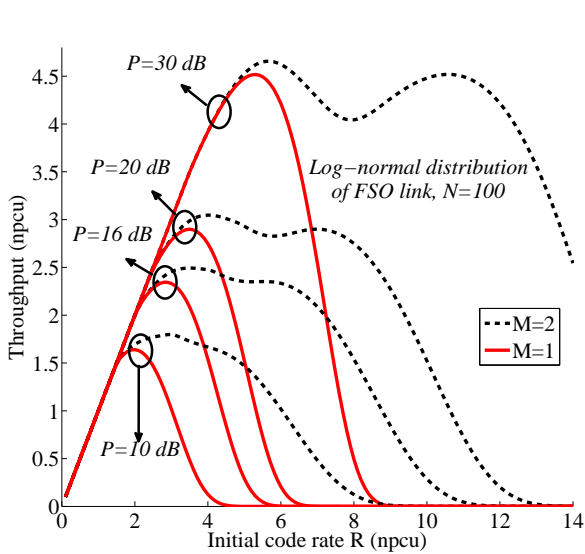}\\ \vspace{-1mm}
\caption{Throughput versus the initial code rate $R$. The FSO link follows a log-normal distribution and $N=100.$} \vspace{-1mm}\label{figure111}
\end{figure}

\begin{figure}
\centering
  \includegraphics[width=0.99\columnwidth]{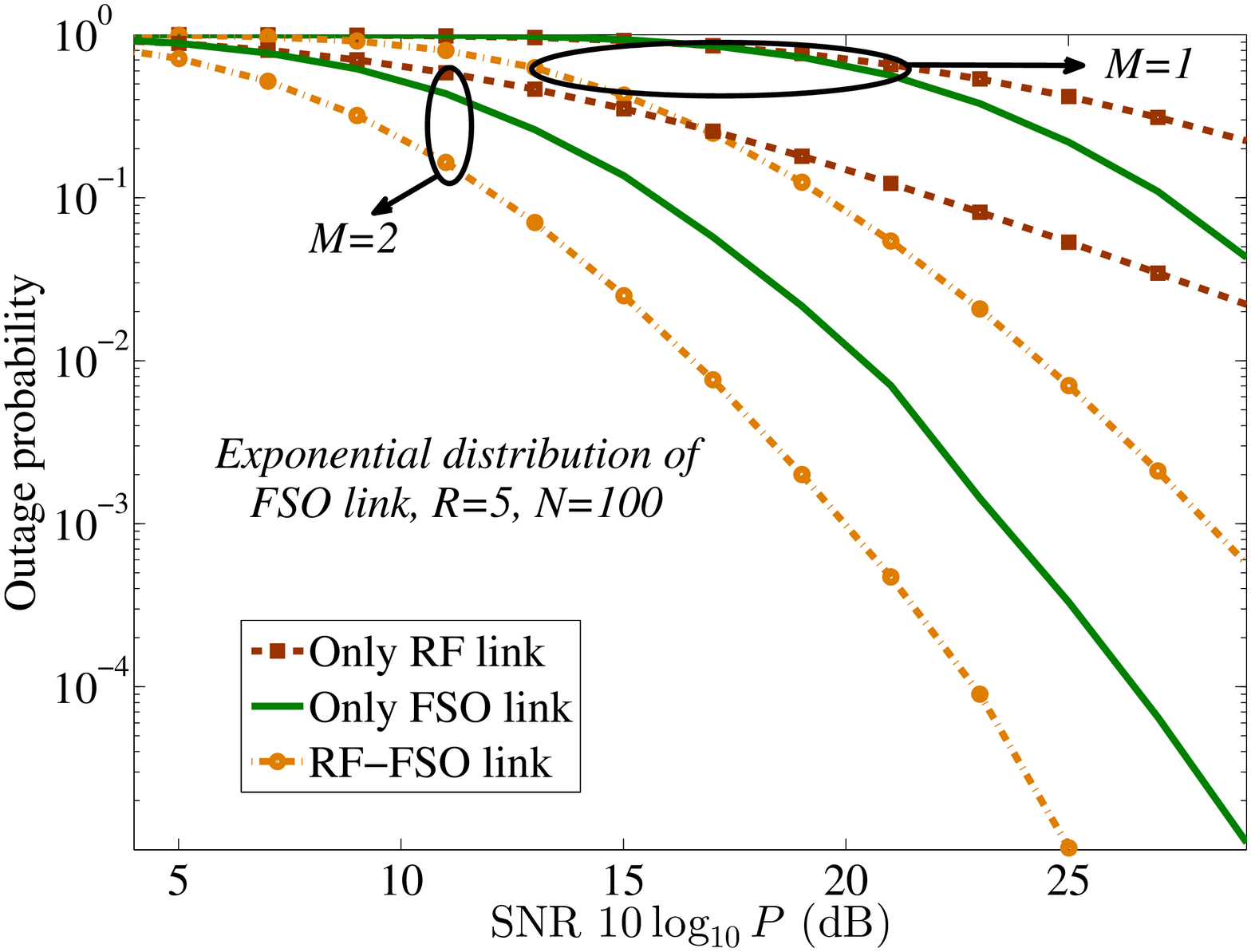}\\ \vspace{-1mm}
\caption{Comparison between the performance of the RF, the FSO, and the RF-FSO based systems. In all cases, the RF and the FSO links are supposed to follow $f_{G_\text{RF}}(x)=\lambda_\text{RF} e^{-\lambda_\text{RF} x}$ and $f_{G_\text{FSO}}(x)=\lambda_\text{FSO} e^{-\lambda_\text{FSO} x}$ where $\lambda_\text{RF}$ and $\lambda_\text{FSO}$ follow normalized log-normal distributions ($R=5$ npcu, $N=100$).} \vspace{-1mm}\label{figure111}
\end{figure}

\emph{On the effect of power allocation:} Figure 11a shows the outage probability for different power allocation schemes and PDFs of the FSO link. Also, Fig. 11b demonstrates the optimal power terms, derived numerically, that minimize the outage probability. Here, the results are presented for the exponential and log-normal PDFs of the FSO link, $N=20, M=2$, $R=5$ npcu, and different sum powers $P=P_\text{RF}+P_\text{FSO} $ (note that the power terms $P, P_\text{RF}$ and $P_\text{FSO}$ are presented in dB). As shown in the figures and in harmony with Lemma 6, adaptive power allocation between the RF and FSO link has marginal effect on the outage probability (Fig. 11a). For small $\lambda_\text{RF}$'s, which is dual to the low SNR performance analysis, the minimum outage probability is achieved by allocating low power to the RF link (see Lemma 7). However, the difference between the optimal power terms of the RF and FSO links decreases with the total transmission power (Fig. 11b). Thus, based on Lemma 6 and Fig. 11 (and also due to complexity of adaptive power allocation), for moderate/high SNRs uniform power allocation is recommended for HARQ-based RF-FSO links. Finally, in harmony with intuitions, Fig. 11b indicates that, although the difference between the optimal power terms of the RF and FSO links is negligible, to minimize the outage probability, higher power should be assigned to the link with better long-term channel quality.

\section{Conclusion}
This paper studied the performance of RF-FSO systems in the cases with perfect CSI at the receivers. Considering different relative coherence times for the RF and FSO links, we derived closed-form expressions for the message decoding probabilities, throughput, and outage probability of the RF-FSO systems using HARQ. The results show that the joint implementation of RF and FSO links leads to considerable throughput and outage probability improvement, compared to the cases utilizing either the RF or the FSO link separately. Moreover, adaptive power allocation improves the performance of RF-FSO based systems, while at high SNRs the optimal, in terms of throughput/outage probability, power allocation converges to uniform power allocation, independently of the links channel conditions. Block error rate analysis of the RF-FSO links in the presence of finite-length codewords is an interesting extension of the work presented in this paper. Here, the results of \cite{5452208} can be of great help.
\vspace{-0mm}

\begin{figure*}
\centering
  \includegraphics[width=1.99\columnwidth]{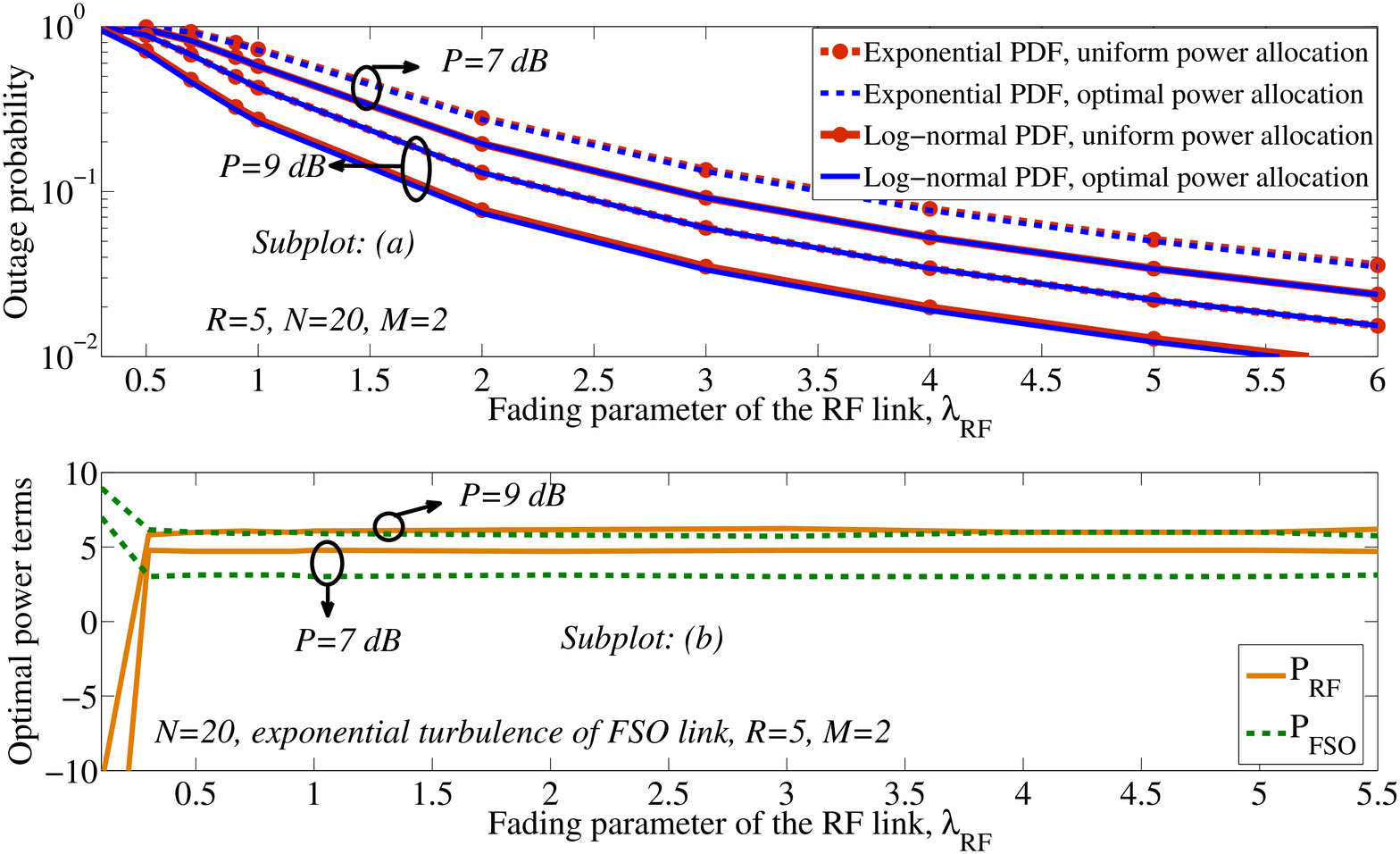}\\ \vspace{-1mm}
\caption{On the effect of power allocation. Subplots (a) and (b) respectively show the outage probability for different power allocation schemes and the optimal power terms minimizing the outage probability. For the exponential and log-normal turbulence of the FSO link, the PDF is given by $f_{G_\text{FSO}}(x)=\lambda_\text{FSO} e^{-\lambda_\text{FSO} x}$ and $f_{G_\text{FSO}}(x)=\frac{1}{\sqrt{2\pi}{\delta x}}e^{-\frac{(\log(x)-\varpi)^2}{2\delta^2}}$ with $\lambda_\text{FSO}=1,$ $\delta=1,$ and $\varpi=0$ ($R=5$ npcu, $M=2$, and $N=20$).} \vspace{-1mm}\label{figure111}
\end{figure*}

\appendix
\subsection{Deriving the Mean and Variance of the Equivalent Gaussian Random variable for Gamma-Gamma PDF of FSO Link}
Using the expansion technique \cite[p. 378]{handbookbessel}
\begin{align}\label{eq:gammagammaapx1}
&K_n(x)= \sqrt{\frac{\pi}{2x}}e^{-x}\left(\sum_{i=0}^\infty{\frac{\alpha_i}{i!(8x)^i}}\right), \nonumber\\&\alpha_i=\prod_{j=1}^i{\left(4n^2-(2j-1)^2\right)},i>0, \alpha_0=1,
\end{align}
and \cite[p. 378]{handbookbessel}
\begin{align}\label{eq:gammagammaapx12}
K_n(x)\simeq \frac{\Gamma(n)}{2}\left(\frac{2}{x}\right)^n, \, n\ne 0, \text{small } x\text{'s}
\end{align}
we write
\begin{align}\label{eq:eqmugamma1}
\mu&=\int_0^\infty{\log(1+P_\text{FSO}x)f_{G_\text{FSO}}(x)\text{d}x}\nonumber\\&
\mathop  \simeq \limits^{(o)}\frac{(ab)^b\Gamma(a-b)}{\Gamma(a)\Gamma(b)}\int_0^\xi{\log(1+P_\text{FSO}x)x^{b-1}\text{d}x}\nonumber\\&+\frac{(ab)^{\frac{a+b}{2}}}{\Gamma(a)\Gamma(b)}\frac{\sqrt{\pi}}{\sqrt[4]{ab}}
\times\nonumber\\&\,\,\,\,\,\,\,\,\,\,\,\int_\xi^\infty{\log(P_\text{FSO}x)x^{\frac{a+b}{2}-\frac{5}{4}}e^{-2\sqrt{abx}}(\sum_{i=0}^{\infty}{\frac{\alpha_i}{i!(16\sqrt{abx})^i}})\text{d}x}
\nonumber\\&
\mathop  = \limits^{(p)}\frac{(ab)^b\Gamma(a-b)}{\Gamma(a)\Gamma(b)}\int_0^\xi{\log(1+P_\text{FSO}x)x^{b-1}\text{d}x}\nonumber\\&
+\frac{(ab)^{\frac{a+b}{2}}}{\Gamma(a)\Gamma(b)}\frac{\sqrt{\pi}}{\sqrt[4]{ab}}\times\nonumber\\&\,\,\,\,\,\,\,\sum_{i=0}^\infty{\frac{\alpha_i}{i!8^i(4ab)^{\frac{a+b}{2}-\frac{1}{4}}}\int_{4ab\xi}^\infty{\log\left(\frac{P_\text{FSO}u}{4ab}\right)u^{\frac{a+b}{2}-\frac{5}{4}-\frac{i}{2}}e^{-\sqrt{u}}\text{d}u}}\nonumber\\&
=\frac{(ab)^b\Gamma(a-b)}{\Gamma(a)\Gamma(b)}\frac{\xi^b}{b(b+1)}\times\nonumber\\&\bigg((b+1)\log(1+P_\text{FSO}\xi)-P_\text{FSO}\xi\prescript{}{2}F_{1}(1,b+1;b+2;-P_\text{FSO}\xi)\bigg)
\nonumber\\&+
\frac{(ab)^{\frac{a+b}{2}}}{\Gamma(a)\Gamma(b)}\frac{\sqrt{\pi}}{\sqrt[4]{ab}}\times\nonumber\\&\,\,\,\,\,\,\,\,\,\,\,\,\,\sum_{i=0}^\infty{\frac{\alpha_i}{i!8^i(4ab)^{\frac{a+b}{2}-\frac{1}{4}}}\bigg(\mathcal{U}(x\to\infty,i)-\mathcal{U}(4ab\xi,i)\bigg)}
,\nonumber\\&
\mathcal{U}(x,i)=\frac{1}{(\frac{a+b}{2}-\frac{1}{4}-\frac{i}{2})^2}\times\nonumber\\&\bigg(2\left(\frac{a+b}{2}-\frac{1}{4}-\frac{i}{2}\right)^2\bigg(\Gamma({a+b}-\frac{1}{2}-{i})\log(x)\nonumber\\&\,\,\,\,\,\,\,\,\,\,\,\,-\log\bigg(\frac{P_\text{FSO}}{4ab}x\bigg)\Gamma\bigg({a+b}-\frac{1}{2}-{i},\sqrt{x}\bigg)\bigg)\nonumber\\&-x^{\frac{a+b}{2}-\frac{1}{4}-\frac{i}{2}}\prescript{}{2}F_{2}\bigg({a+b}-\frac{1}{2}-{i},{a+b}-\frac{1}{2}-{i}\nonumber\\&\,\,\,\,\,\,\,\,\,\,\,\,\,\,\,\,\,\,\,\,\,\,\,\,\,\,\,\,\,\,\,\,\,\,\,\,\,\,\,\,\,\,\,\,\,\,\,\,\,\,;{a+b}+\frac{1}{2}-{i},{a+b}+\frac{1}{2}-{i};-\sqrt{x}\bigg)\bigg).
\end{align}
Here, $(o)$ is obtained by two integration parts where in the first (resp. second) integration we use the approximation (\ref{eq:gammagammaapx12}) (resp. $\log(1+x)\simeq \log(x)$ and (\ref{eq:gammagammaapx1})) for small (resp. large) values of $x$. Then, $(p)$ comes from the variable transform  $4abx=u$ and the last equality follows from some manipulations and the definition of Gamma incomplete function and the generalized hypergeometric function $\prescript{}{a_1}F_{a_2}(.).$

Finally, using (\ref{eq:eqmugamma1}), the variance of the equivalent Gaussian variance in Gamma-Gamma distribution of the FSO link is given by $\sigma^2=\rho^2-\mu^2$ where
\begin{align}
\rho^2&=\int_0^\infty{\log^2(1+P_\text{FSO}x)f_{G_\text{FSO}}(x)\text{d}x}\nonumber\\&
\simeq \frac{(ab)^b\Gamma(a-b)}{\Gamma(a)\Gamma(b)}\int_0^\xi{\log^2(1+P_\text{FSO}x)x^{b-1}\text{d}x}
\nonumber\\&+\frac{(ab)^{\frac{a+b}{2}}}{\Gamma(a)\Gamma(b)}\frac{\sqrt{\pi}}{\sqrt[4]{ab}}\times\nonumber\\&\,\,\,\,\,\,\,\,\,\,\,\int_\xi^\infty{\log^2(P_\text{FSO}x)x^{\frac{a+b}{2}-\frac{5}{4}}e^{-2\sqrt{abx}}(\sum_{i=0}^{\infty}{\frac{\alpha_i}{i!(16\sqrt{abx})^i}})\text{d}x}
\nonumber\\&
=\frac{(ab)^b\Gamma(a-b)}{\Gamma(a)\Gamma(b)}\int_0^\xi{\log^2(1+P_\text{FSO}x)x^{b-1}\text{d}x}\nonumber\\&
+\frac{(ab)^{\frac{a+b}{2}}}{\Gamma(a)\Gamma(b)}\frac{\sqrt{\pi}}{\sqrt[4]{ab}}\times\nonumber\\&\,\,\,\,\,\,\,\sum_{i=0}^\infty{\frac{\alpha_i}{i!8^i(4ab)^{\frac{a+b}{2}-\frac{1}{4}}}\int_{4ab\xi}^\infty{\log^2(\frac{P_\text{FSO}u}{4ab})u^{\frac{a+b}{2}-\frac{5}{4}-\frac{i}{2}}e^{-\sqrt{u}}\text{d}u}}\nonumber\\&
=\frac{(ab)^b\Gamma(a-b)}{\Gamma(a)\Gamma(b)} (\mathcal{J}(\xi)-\mathcal{J}(0))
\nonumber\\&+
\frac{(ab)^{\frac{a+b}{2}}}{\Gamma(a)\Gamma(b)}\frac{\sqrt{\pi}}{\sqrt[4]{ab}}\times\nonumber
\end{align}
\begin{align}\label{eq:eqsigmagamma1}
&\sum_{i=0}^\infty{\frac{\alpha_i}{i!8^i(4ab)^{\frac{a+b}{2}-\frac{1}{4}}}\left(\mathcal{H}(4ab\xi,i)-\mathcal{H}(x\to\infty,i)\right)}
,\nonumber\\&
\mathcal{J}(x)=\frac{1}{b(-P_\text{FSO})^b}\times\nonumber\\&\bigg(-2b(1+P_\text{FSO}x)\prescript{}{4}F_{3}(1,1,1,1-b;2,2,2;1+P_\text{FSO}x)\nonumber\\&+2b(1+P_\text{FSO}x)\log(1+P_\text{FSO}x)\times\nonumber\\&\prescript{}{3}F_{2}(1,1,1-b;2,2;1+P_\text{FSO}x)\nonumber\\&\,\,\,\,\,\,\,\,\,\,\,\,\,\,\,\,\,\,\,\,\,\,\,\,\,\,\,\,\,\,\,\,\,\,\,\,\,\,+((-P_\text{FSO}x)^b-1)\log^2(1+P_\text{FSO}x)\bigg),
\nonumber\\&
\mathcal{H}(x,i)=\frac{2}{(\frac{a+b}{2}-\frac{1}{4}-\frac{i}{2})^3}\times\nonumber\\&\bigg(\left(\frac{a+b}{2}-\frac{1}{4}-\frac{i}{2}\right)\times\nonumber\\&\bigg(x^{\frac{a+b}{2}-\frac{1}{4}-\frac{i}{2}}\log\left(\frac{P_\text{FSO}}{4ab}x\right) \prescript{}{2}F_{2}\bigg({a+b}-\frac{1}{2}-{i},\nonumber\\&{a+b}-\frac{1}{2}-{i};{a+b}+\frac{1}{2}-{i},{a+b}+\frac{1}{2}-{i};-\sqrt{x}\bigg)\nonumber\\&+\left(\frac{a+b}{2}-\frac{1}{4}-\frac{i}{2}\right)^2\bigg(\Gamma\left({a+b}-\frac{1}{2}-{i}\right)\times\nonumber\\&\,\,\,\,\,\,\,\,\,\,\,\,\,\,\,\,\,\,\,\,\,\,\,\,\,\,\,\,\,\,\,\,\,\,\,\,\log(x)\bigg(\log(x)-2\log\left(\frac{P_\text{FSO}}{4ab}x\right)\bigg)\nonumber\\&+\log^2(\frac{P_\text{FSO}}{4ab}x)\Gamma\left({a+b}-\frac{1}{2}-{i},\sqrt{x}\right)\bigg)\bigg)-x^{\frac{a+b}{2}-\frac{1}{4}-\frac{i}{2}}\times\nonumber\\&\prescript{}{3}F_{3}\bigg({a+b}-\frac{1}{2}-{i},{a+b}-\frac{1}{2}-{i},{a+b}-\frac{1}{2}-{i}\nonumber\\&;{a+b}+\frac{1}{2}-{i},{a+b}+\frac{1}{2}-{i},{a+b}+\frac{1}{2}-{i};-\sqrt{x}\bigg)\bigg)
\end{align}
with the (in)equalities following the same procedure as in (\ref{eq:eqmugamma1}). Finally, note that in (\ref{eq:eqmugamma1}) and (\ref{eq:eqsigmagamma1}) the approximation is tight for different values of $\xi\ge 0.$ Then, the appropriate value of $\xi$ is determined numerically such that the difference between the exact and the approximate probabilities is minimized.

\vspace{-0mm}
\bibliographystyle{IEEEtran} 
\bibliography{masterFSO2}

\vfill
\end{document}